\documentclass{aa}
\usepackage{txfonts}
\usepackage{epsfig}
\usepackage{subfigure}

\newcommand{\gapprox}{\hbox{\lower .8ex\hbox{$\,\buildrel > \over\sim\,$}}}
\newcommand{\lapprox}{\hbox{\lower .8ex\hbox{$\,\buildrel < \over\sim\,$}}}
\usepackage{times}
\usepackage{natbib}
\usepackage{rotating}
\bibpunct{(}{)}{;}{a}{}{,}

\begin{document}

\title{An XMM-Newton spatially-resolved study of 
metal abundance evolution in distant galaxy clusters}

\author{A. Baldi\inst{1,2} 
\and S. Ettori\inst{2,3} \and S. Molendi\inst{4} \and
I. Balestra\inst{5} \and F. Gastaldello\inst{4,6} \and P. Tozzi\inst{7,8}}

\offprints{A. Baldi, \email{alessandro.baldi@oabo.inaf.it}}

\institute{Dipartimento di Astronomia, Universit\`a di Bologna, via Ranzani 1, I--40127, Bologna, Italy 
\and INAF, Osservatorio Astronomico di Bologna, via Ranzani 1, I--40127, Bologna, Italy
\and INFN, Sezione di Bologna, viale Berti Pichat 6/2, I--40127 Bologna, Italy
\and INAF-IASF, via Bassini 15, I--20133, Milan, Italy
\and Max-Planck-Institut f\"ur Extraterrestrische Physik, Postfach 1312, 85741 Garching, Germany 
\and Department of Physics and Astronomy, University of California at Irvine, 4129 Frederick Reines Hall, Irvine, CA 92697--4575, U.S.A.
\and Osservatorio Astronomico di Trieste, via G.B. Tiepolo 11, I--34131, Trieste, Italy
\and INFN, National Institute for Nuclear Physics, Trieste, Italy}

\date{Received 2011}

\titlerunning{An XMM-Newton study of metal abundance in distant galaxy clusters}
\authorrunning{A. Baldi et al.}

\abstract {We present an XMM-Newton analysis 
of the X-ray spectra of 39 clusters of galaxies at $0.4<z<1.4$, covering a temperature range of
$1.5\,\la kT\,\la 11$~keV.} {The main goal of this paper is to study how the abundance
evolves with redshift not only by means of a single emission measure performed on the whole cluster 
but also by spatially resolving the cluster emission.} 
{We performed a spatially resolved spectral analysis, using Cash statistics and modeling the XMM-Newton background instead of
subtracting it, by analyzing the contribution of the core emission to the observed metallicity.} 
{We do not observe a statistically significant ($>2\sigma$) 
abundance evolution with redshift. The most
significant deviation from no evolution (at a 90\% confidence level) is observed by
considering the emission from the whole cluster ($r<0.6r_{500}$), which can be parametrized as $Z\propto (1+z)^{-0.8\pm0.5}$.
Dividing the emission into three radial bins, no significant evidence of abundance evolution is observed when fitting
the data with a power law. We find close agreement with measurements presented in previous studies.
Computing the error-weighted mean of the spatially resolved abundances into three redshift bins, we find that it is consistent with
being constant with redshift. Although the large error bars in the measurement of the weighted-mean abundance prevent us from 
claiming any statistically significant spatially resolved evolution, the trend with $z$ in the 0.15-0.4$r_{500}$ radial bin 
complements nicely the measures of Maughan et al., and broadly agrees with theoretical predictions.
We also find that the data points derived from the spatially resolved analysis are well-fitted by the relation
$Z(r,z)=Z_0(1+(r/0.15r_{500})^2)^{-a}((1+z)/1.6)^{-\gamma}$, where $Z_0=0.36\pm0.03$, 
$a=0.32\pm0.07$, and $\gamma=0.25\pm0.57$, which represents a significant negative trend 
of $Z$ with radius and no significant evolution with redshift.}
{We present the first attempt to determine the evolution of abundance at different positions in the clusters and with
redshift. However, the sample size and the low-quality data statistics associated with most of the clusters studied prevents us from 
drawing any statistically significant conclusion about the different evolutionary path that the different regions of the
clusters may have traversed.}


\maketitle

\section{Introduction}

Clusters of galaxies are the largest virialized structures in the
Universe, arising from the gravitational collapse of high peaks of
primordial density perturbations. They represent unique signposts
where the physical properties of the cosmic diffuse baryons can be
studied in great detail and used to trace the past history of cosmic
structure formation \citep[e.g.][]{peebles93, coles95, peacock99, 
rosati02, voit05}.
The gravitational potential well of clusters is permeated by a hot
thin gas, which is typically enriched with metals ejected form
supernovae (SNe) explosions through subsequent episodes of star
formation \citep[e.g.][]{matteucci88,renzini97}.
This gas reaches temperatures of several $10^7$ K and therefore emits mainly via
thermal bremsstrahlung and is easily detectable in the X-rays.
Although the amount of energy supplied to the
intra-cluster medium (ICM) by SNe explosions depends on several
factors (e.g. the physical condition of the ICM at the epoch of the
enrichment) and cannot be obtained directly from X--ray observations,
the radial distribution of metals in the ICM, as well as their abundance as a
function of time, represent the `footprint' of cosmic
star formation history and are crucial to trace the effect of SN feedback 
on the ICM at different cosmic epochs\citep[e.g.][]{ettori05,borgani08}.

Several studies of the radial distributions of metals in the ICM
at low redshift are present in the literature \citep[e.g.][]{finoguenov00a,degrandi01,irwin01,
tamura04,vikhlinin05,baldi07,leccardi08b}.
On the other hand just a handful of works have tackled the evolution of metal
abundance in the ICM at larger look-back times.
Measurements of the metal content of the ICM
at high$-z$ has been obtained with single emission-weighted estimates from Chandra
and XMM-Newton exposures of 56 clusters at $0.3\la z\la 1.3$ in \citet{balestra07} (hereafter BLS07).
They measured the iron abundance within (0.15-0.3)$R_{vir}$ and found
a negative evolution of $Z_{Fe}$ with redshift, for clusters at $z\ga 0.5$ that have
a constant average Fe abundance of $\approx0.25Z_\odot$, while objects
in the redshift range $0.3\la z \la0.5$ have a $Z_{Fe}$ that is significantly higher
($\approx0.4Z_\odot$). 
This result was confirmed by \citet{maughan08} (hereafter MAU08) for a sample of 116 Chandra clusters
at $0.1<z<1.3$. Stacking the cluster spectra extracted within $r_{500}$ 
in different redshift bins, they found that the metal abundance in the ICM drops by a factor of $\sim50\%$
between clusters in the local Universe and clusters at $z=1$. This evolution is still present 
when the cluster cores ($r<0.15r_{500}$) are excluded from the abundance measurements, 
indicating that the evolution is not simply driven by the disappearance of relaxed, cool-core clusters 
(which are known to have enhanced core metal abundances) from the population at $z\gtrsim 0.5$ \citep{vikhlinin07}.
\citet{anderson09} found a similar drop in abundance between $z=0.1$ and $z\sim1$, by adding the data from the 
MAU08 sample and from a low redshift ($z<0.3$) XMM-Newton sample \citep{snowden08}, to an additional 
sample of 29 galaxy clusters at $0.3<z<1.3$ observed by XMM-Newton.

In this work, we aim to study the evolution of metal abundance with redshift by also considering
its radial dependence (using two or three spatial bins) not only to disentangle the evolution of the
cores from the rest of the clusters but also to study the evolution in the outer regions ($r>0.4r_{500}$).
The plan of the paper is the following. 
In \S~2, we describe the sample and the data
reduction procedure. 
We describe our spectral analysis strategy in \S~3 with particular
focus on the background treatment procedure (\S~3.3).
In \S~4, we present the results obtained in our analysis 
mainly regarding abundance evolution with redshift.
We discuss our results in \S~5 and summarize our conclusions in
\S~6. 

We adopt a cosmological model with $H_0=70$ km/s/Mpc,
$\Omega_m=0.3$, and $\Omega_\Lambda=0.7$ throughout. Confidence
intervals are quoted at $1\sigma$ unless otherwise stated.

\section{Sample selection and data reduction}

\begin{figure}
\includegraphics[width=9.0 cm, angle=0]{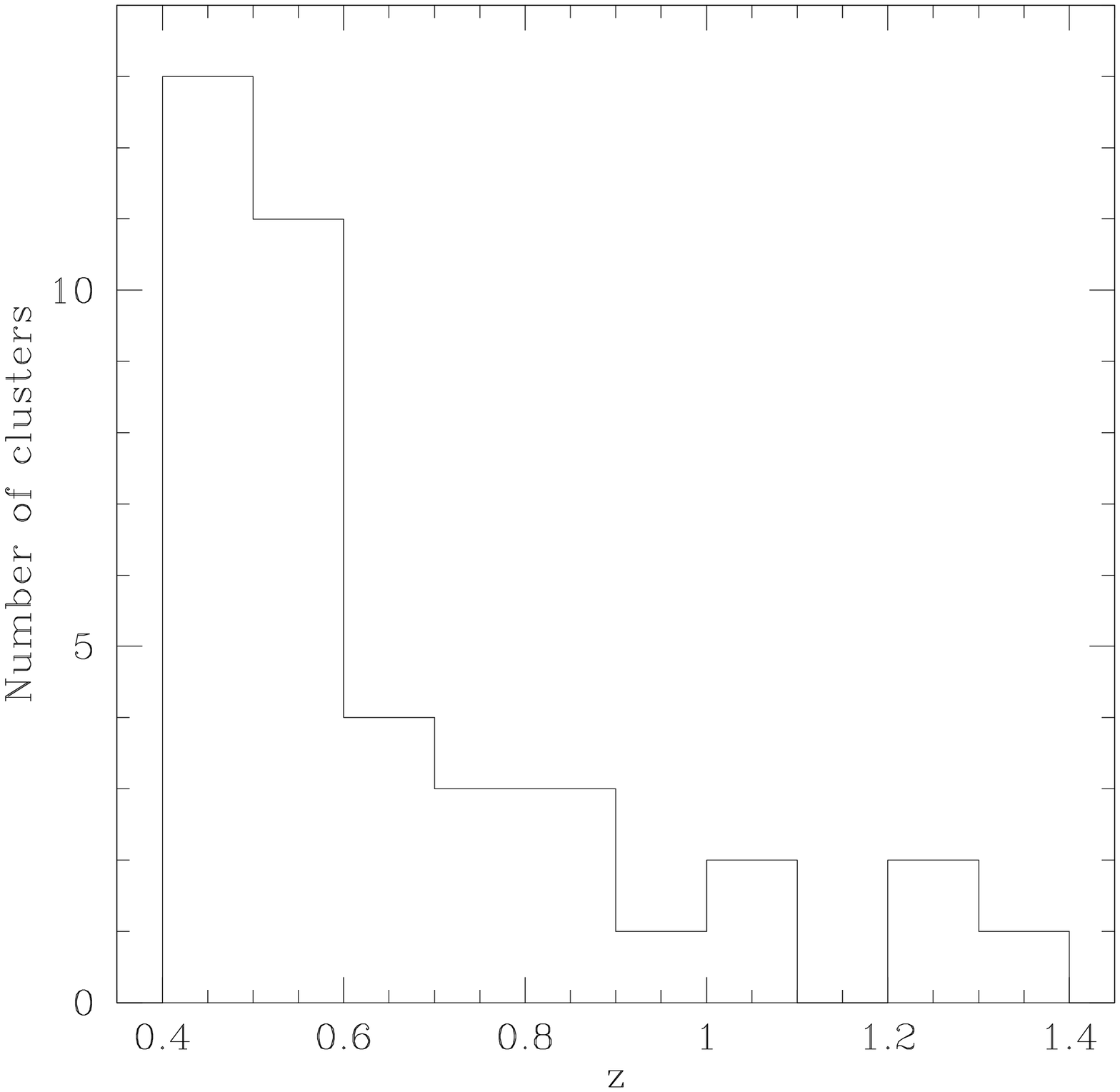}
\caption{Redshift distribution of the XMM-Newton galaxy cluster sample.}
\label{histoz}
\end{figure}

In Table~\ref{exposures}, we present the list of galaxy clusters
analyzed in this paper. The selected clusters are all the clusters at 
$z>0.4$ observed at least once by XMM-Newton and with sufficient signal-to-noise ratio (S/N)
to allow a spatially resolved spectral analysis (at least in two spatial bins).
To be included in the analysis, each bin was required to have a S/N larger than 60\% and to have at least 300 MOS net counts.
This was a trade-off in order to have both a sizeable sample and to avoid the introduction of excessive noise in 
the measure of the abundance.
Figure~\ref{histoz} shows the redshift distribution of the sample, which is clearly peaked at low
redshifts with more than half of the sample being at $z<0.6$.

The observation data files (ODF) were processed to produce calibrated event 
files using the XMM-{\em Newton} Science Analysis System ({\tt SAS v9.0.0})
processing task {\tt EPPROC} and {\tt EMPROC} for the pn and MOS, respectively. 
Unwanted hot, dead, or flickering pixels were removed as were events due to 
electronic noise. However, the pn data were only used in the imaging analysis,
because of the possible effects of an under- or over-estimation of the pn
particle background on the analysis of the extended sources 
\citep[see e.g.][and XMM-ESAS cookbook\footnote{ftp://xmm.esac.esa.int/pub/xmm-esas/xmm-esas.pdf}]{leccardi08a}.
As we also tested on a representative subsample of our clusters (in redshift, flux, and temperature), this
leads to inconsistencies with the MOS detectors in the measurements of temperature and abundance. 
On the basis of the systematic differences between MOS and pn, a joint fit of pn spectra with MOS spectra 
would be unjustified, since it would introduce both a bias in the best-fit values and artificially smaller 
errors (caused by the worsened spectral fits).

To search for periods of high background flaring, light curves for pattern=0 events 
and for pattern$\leq 4$ in the $10-12$~keV are produced for 
the pn and the two MOS, respectively. 
The soft proton cleaning was performed using a double-filtering process. We extracted a light curve in 100s bins in 
the 10-12 keV energy band by excluding the central CCD, applied a threshold of 0.20 cts s$^{-1}$, produced a GTI file, and 
generated the filtered event file accordingly. This first step allows most flares to be eliminated, although softer 
flares may exist such that their contribution above 10 keV is negligible. We then extracted a light curve in the 2-5 keV 
band and fit the histogram obtained from this curve with a Gaussian distribution. Since most flares had been rejected 
in the previous step, the fit was usually very good. We calculated both the mean count rate, $\mu$, and the standard deviation, 
$\sigma$, applied a threshold of $\mu+3\sigma$ to the distribution, and generated the filtered event file. The intervals 
of very high background were removed using a 3$\sigma$ clipping algorithm
and the light-curves were then visually inspected to remove the background flaring periods
not detected by the algorithm.  
In Table \ref{exposures}, we list the resulting clean exposure times for the pn 
and the MOS detectors.

We selected events with patterns 0 to 4 (single and double) for the pn and 
with patterns 0 to 12 (single, double, and quadruple) for the MOS. Further filtering was applied
to remove events with low spectral quality. For the pn in particular, we removed the
out-of-field events ((FLAG \& 0x10000) = 0)), and the events close to bad pixels((FLAG \& 0x20) = 0), 
on offset columns ((FLAG \& 0x8) = 0), and close to a CCD window ((FLAG \& 0x4) = 0).
We merged the event files from each detector (and in some cases from multiple observations) to create a single event file.
The total event file was used to extract a 0.5-8 keV band image of each cluster, 
but not for the spectral analysis where the spectra
were extracted from each individual detector (and observation) and fitted simultaneously (See \S~\ref{specan} below).
The use of the $pn$ data in the resulting image was motivated by the possibility of detecting the point sources to remove from
the cluster spectra at a lower flux limit. A list of point-like sources was created using the {\tt EBOXDETECT} task and then visually
inspected to check for false detections. When one or more point sources were located inside an extraction region, 
circular regions of radius $\ga15''$ (depending on the source brightness), centered at the position of each point source,
were excised from the spectral extraction.



\begin{table*}
\caption{XMM-{\em Newton} archival 
observations of distant clusters of galaxies.\label{exposures}}
\centering
\begin{tabular}{lcrrccccl}
\hline
Cluster & $z$ & Obs. Date & Obs. ID & $T_{clean,MOS1}$ & $T_{clean,MOS2}$ & $T_{clean,pn}$ & $N_H$ & Radial bins\tablefootmark{a} \\
 & & & & (ksec) & (ksec) & (ksec) & (10$^{20}$ cm$^{-2}$) & \\
\hline\hline
A851              &  0.407   & 2000 Nov 06 & 0106460101 & 41.7 & 41.1 & 30.3 & 1.0 & 1,2,3 \\
\hline
RXJ1213.5+0253    &  0.409   & 2001 Dec 30 & 0081340801 & 21.8 & 22.0 & 14.3 & 1.8 & 2 \\
\hline
RXCJ0856.1+3756   &  0.411   & 2005 Oct 10 & 0302581801 & 24.2 & 23.9 & 14.7 & 3.2 & 1,2,3 \\
\hline
RXJ2228.6+2037    &  0.412   & 2003 Nov 18 & 0147890101 & 24.6 & 24.4 & 19.3 & 4.3 & 1,2,3 \\
\hline
RXCJ1003.0+3254   &  0.416   & 2005 Nov 03 & 0302581601 & 25.0 & 25.1 & 15.7 & 1.7 & 1,2,3 \\
\hline
MS0302.5+1717     &  0.425   & 2002 Aug 23 & 0112190101 & 12.5 & 12.7 &  7.3 & 9.4 & 1,2 \\
\hline
RXCJ1206.2-0848   &  0.440   & 2007 Dec 09 & 0502430401 & 29.1 & 28.8 & 20.6 & 3.7 & 1,2,3 \\
\hline
IRAS09104+4109    &  0.442   & 2003 Apr 27 & 0147671001 & 12.1 & 12.3 & 8.3 & 1.4 & 1,2,3 \\
\hline
RXJ0221.1+1958    &  0.450   & 2005 Jul 14 & 0302581301 &  5.4 &  6.5 &  1.2 & 9.1 & 1,2 \\
\hline
RXJ1347.5-1145    &  0.451   & 2002 Jul 31 & 0112960101 & 32.6 & 32.4 & 27.5 & 4.9 & 1,2,3 \\
\hline
RXJ1311.5-0551    &  0.461   & 2006 Jan 17 & 0302582201 & 25.1 & 26.0 & 19.1 & 2.4 & 1,2,3 \\
                  &          & 2007 Jul 26 & 0502430101 & 34.6 & 36.0 & 21.2 &  &  \\
\hline
RXJ0522.2-3625    &  0.472   & 2005 Aug 14 & 0302580901 & 19.2 & 19.4 & 16.0 & 3.6 & 2 \\
\hline
RXJ2359.5-3211    &  0.478   & 2005 Jun 13 & 0302580501 & 38.8 & 38.5 & 29.5 & 1.2 & 1,2 \\
\hline
CLJ0030+2618      &  0.500   & 2005 Jul 06 & 0302581101 & 15.0 & 14.0 & - & 3.7 & 1,2,3 \\
                  &          & 2006 Jul 27 & 0402750201 & 27.0 & 27.4 & 19.6 &  &  \\
                  &          & 2006 Dec 19 & 0402750601 & 28.5 & 28.6 & 20.6 &  &  \\
\hline
WARPJ2146.0+0423  &  0.531   & 2005 Jun 10 & 0302580701 & 20.7 & 20.8 & 17.7 & 4.8 & 2 \\
\hline
RXJ0018.8+1602    &  0.541   & 2007 Dec 14 & 0502860101 & 41.1 & 42.7 & 30.1 & 3.8 & 1,2 \\
\hline
MS0015.9+1609     &  0.541   & 2000 Dec 29 & 0111000101 & 30.6 & 30.1 & 20.4 & 4.0 & 1,2,3 \\
                  &          & 2000 Dec 30 & 0111000201 &  5.5 &  5.4 &  - &  &  \\
\hline
WARPJ0848.8+4456  &  0.543   & 2001 Oct 15 & 0085150101 & 39.1 & 39.6 & 33.7 & 2.8 & 2 \\
                  &          & 2001 Oct 21 & 0085150201 & 23.3 & 25.2 & 13.0 &  &  \\
                  &          & 2001 Oct 21 & 0085150301 & 25.2 & 26.4 & 17.9 &  &  \\
\hline
CLJ1354-0221      &  0.546   & 2002 Jul 19 & 0112250101 &  6.8 &  7.7 & 0.8 & 3.2 & 2 \\
\hline
WARPJ1419.9+0634E &  0.549   & 2005 Jul 07 & 0303670101 & 43.1 & 43.4 & 29.9 & 2.2 & 1,2 \\
\hline
MS0451.6-0305     &  0.550   & 2004 Sep 16 & 0205670101 & 25.5 & 26.1 & 19.4 & 3.9 & 1,2,3 \\
\hline
RXJ0018.3+1618    &  0.551   & 2000 Dec 29 & 0111000101 & 30.5 & 29.6 & 22.4 & 3.9 & 2 \\
\hline
WARPJ1419.3+0638  &  0.574   & 2005 Jul 07 & 0303670101 & 43.1 & 43.4 & 29.9 & 2.2 & 1,2,3 \\
\hline
MS2053.7-0449     &  0.583   & 2001 Nov 14 & 0112190601 & 16.4 & 16.5 &  8.9 & 4.6 & 2 \\
\hline
MACSJ0647.7+7015  &  0.591   & 2008 Oct 09 & 0551850401 & 52.9 & 53.7 & 32.3 & 5.4 & 1,2,3 \\
                  &          & 2009 Mar 04 & 0551851301 & 33.2 & 34.6 & 18.4 &  &  \\
\hline
CLJ1120+4318      &  0.600   & 2001 May 08 & 0107860201 & 18.3 & 18.7 & 13.6 & 3.0 & 1,2,3 \\
\hline
CLJ1334+5031      &  0.620   & 2001 Jun 07 & 0111160101 & 36.1 & 36.5 & 27.0 & 1.1 & 1,2,3 \\
\hline
MACSJ0744.9+3927  &  0.698   & 2008 Oct 17 & 0551850101 & 40.4 & 41.4 & 22.4 & 5.7 & 1,2,3 \\
                  &          & 2009 Mar 21 & 0551851201 & 63.8 & 67.0 & 35.8 &  &  \\
\hline
WARPJ1342.8+4028  &  0.699   & 2002 Jun 08 & 0070340701 & 33.3 & 33.6 & 25.0 & 0.8 & 2,3 \\
\hline
CLJ1103.6+3555    &  0.780   & 2001 May 15 & 0070340301 &  7.1 &  7.4 &  4.4 & 2.6 & 2 \\
                  &          & 2004 May 11 & 0205370101 & 35.2 & 36.0 & 27.4 &  &  \\
\hline
MS1137.5+6625     &  0.782   & 2000 Oct 05 & 0094800201 & 12.8 & 16.2 &  4.9 & 1.0 & 1,2,3 \\
\hline
CLJ1216.8-1201    &  0.794   & 2003 Jul 06 & 0143210801 & 24.0 & 24.2 & 19.7 & 3.3 & 2,3 \\
\hline
MS1054.4-0321     &  0.823   & 2001 Jun 21 & 0094800101 & 22.7 & 23.6 & 19.1 & 3.1 & 1,2,3 \\
\hline
CLJ0152.7-1357    &  0.831   & 2002 Dec 24 & 0109540101 & 50.6 & 50.9 & 40.3 & 1.3 & 1,2,3 \\
\hline
CLJ1226.9+3332    &  0.890   & 2001 Jun 18 & 0070340501 & 10.3 & 10.7 &  5.3 & 1.8 & 1,2,3 \\
                  &          & 2004 Jun 02 & 0200340101 & 67.2 & 67.5 & 53.4 &  &  \\
\hline
CLJ1429.0+4241    &  0.920   & 2006 Jan 08 & 0300140101 & 33.1 & 33.7 & 18.3 & 1.1 & 1,2 \\
\hline
XLSSC029          &  1.050   & 2005 Jan 01 & 0210490101 & 82.0 & 83.2 & 64.3 & 2.3 & 2,3 \\
\hline
RDCSJ1252-2927    &  1.237   & 2003 Jan 03 & 0057740301 & 66.1 & 66.2 & 54.9 & 6.1 & 2,3 \\
                  &          & 2003 Jan 11 & 0057740401 & 65.4 & 66.7 & 56.1 &  &  \\
\hline
1WGAJ2235.3-2557  &  1.393   & 2006 May 03 & 0311190101 & 76.0 & 76.9 & 61.1 & 1.5 & 2 \\
\hline
\end{tabular}
\tablefoot{
\tablefoottext{a}{Radial bins considered in the spatially resolved spectral analysis for each cluster: 1 = ($0-0.15r_{500}$),
2 = ($0.15-0.4r_{500}$), 3 = ($>0.4r_{500}$).}
}\end{table*}

\section{Spectral analysis strategy}\label{specan}

Our strategy for performing a spatially resolved spectral analysis involves the determination
of the overdensity radius $r_{500}$. This is necessary because we aim to study the cluster abundances and temperatures 
in annuli with inner and outer radii depending on the physical properties of the cluster instead
of the number of counts in each annulus.
We analyze each cluster into two spatial bins: the region of the cluster core (corresponding
to $0<r<0.15r_{500}$) and the region immediately surrounding the core ($0.15r_{500}<r<0.4r_{500}$).
For most clusters, a third bin at $r>0.4r_{500}$, considering the outskirts of the cluster,
could also be analyzed. The spatial extension of the third bin was determined by considering concentric 
annuli of thickness $0.1r_{500}$ at increasing radii ($0.4r_{500}<r<0.5r_{500}$, $0.5r_{500}<r<0.6r_{500}$,
etc.). We computed the source and background counts in each annulus moving towards the cluster outskirts, 
and we ceased to add annuli to the third bin spectrum when the source counts in the annulus fell below 30\% of the total counts. 
We decided to include in the analysis only the spectra with at least 300 net counts, in order to avoid the introduction of
excessive noise in the measure of the abundance.
In Table~\ref{exposures}, the radial bins considered in the spatially resolved spectral analysis are indicated. We have a total
of 26, 39, and 24 spectra in the $0-0.15r_{500}$, $0.15-0.4r_{500}$, and $>0.4r_{500}$ radial bin, respectively.\\
\begin{table*}
\caption{X-ray global properties.\label{xrayprop}}
\centering
\begin{tabular}{lccccc}
\hline
Cluster & $\langle T \rangle$ & $\langle Z \rangle$ & $r_{500}$ & $T_{0-0.6r_{500}}$ & $Z_{0-0.6r_{500}}$ \\
        & (keV) & (Z$_\odot$) & (Mpc) & (keV) & (Z$_\odot$) \\
\hline\hline
A851              &  $5.66\pm0.27$ & $0.26_{-0.06}^{+0.07}$ & 0.975 &  $5.73_{-0.26}^{+0.24}$ &  $0.24\pm0.06$   \\
RXJ1213.5+0253    &  $1.81_{-0.42}^{+0.57}$ & $<0.16$ & 0.532 &   $2.90_{-0.79}^{+0.54}$ &  $0.25_{-0.25}^{+0.27}$  \\
RXCJ0856.1+3756   &  $6.23\pm0.57$ & $0.42\pm0.11$ & 1.023 &  $6.43_{-0.45}^{+0.46}$ &  $0.37\pm0.09$    \\
RXJ2228.6+2037    &  $7.60\pm0.42$ & $0.28\pm0.06$ & 1.137 &  $7.89_{-0.37}^{+0.35}$ &  $0.28\pm0.05$   \\
RXCJ1003.0+3254   &  $2.96_{-0.23}^{+0.36}$ & $0.24_{-0.13}^{+0.14}$ & 0.689 &  $3.27_{-0.18}^{+0.19}$  & $0.52\pm0.12$   \\
MS0302.5+1717     &  $3.97_{-0.64}^{+0.63}$ & $0.04_{-0.16}^{+0.16}$ & 0.799 &  $4.36_{-0.51}^{+0.58}$ &  $0.17_{-0.12}^{+0.15}$   \\
RXCJ1206.2-0848   & $11.01_{-0.65}^{+0.60}$ & $0.26\pm0.06$ & 1.361 &  $10.48_{-0.41}^{+0.40}$ &  $0.27\pm0.04$   \\
IRAS09104+4109    &  $4.91_{-0.42}^{+0.43}$ & $0.41_{-0.12}^{+0.13}$ & 0.886 &   $5.49_{-0.20}^{+0.21}$ &  $0.53\pm0.08$   \\
RXJ0221.1+1958    &  $6.97_{-2.04}^{+1.87}$ & $<0.29$ & 1.062 &   $5.96_{-2.02}^{+1.89}$ &  $0.01_{-0.01}^{+0.23}$   \\
RXJ1347.5-1145    &  $11.23\pm0.43$ & $0.23\pm0.05$ & 1.366 &   $11.11_{-0.24}^{+0.25}$ &  $0.31\pm0.03$  \\
RXJ1311.5-0551    &  $3.15\pm0.36$ & $0.22_{-0.14}^{+0.15}$ & 0.692 &   $3.10\pm0.31$ &  $0.20_{-0.11}^{+0.14}$    \\
RXJ0522.2-3625    &  $3.28_{-0.51}^{+0.62}$ & $0.14_{-0.14}^{+0.23}$ & 0.703 &   $3.88_{-0.47}^{+0.53}$ &  $0.31_{-0.16}^{+0.22}$  \\
RXJ2359.5-3211    &  $3.35_{-0.44}^{+0.54}$ & $0.41_{-0.22}^{+0.26}$ & 0.708 &   $3.26_{-0.35}^{+0.39}$ &  $0.31_{-0.15}^{+0.18}$  \\
CLJ0030+2618      &  $4.69_{-0.30}^{+0.36}$ & $0.44_{-0.10}^{+0.12}$ & 0.836 &   $5.01_{-0.32}^{+0.30}$ &  $0.43\pm0.09$  \\
WARPJ2146.0+0423  &  $3.64_{-0.44}^{+0.51}$ & $0.47_{-0.20}^{+0.28}$ & 0.717 &   $3.40_{-0.34}^{+0.40}$ &  $0.50_{-0.18}^{+0.24}$  \\
RXJ0018.8+1602    &  $3.69_{-0.46}^{+0.56}$ & $0.45_{-0.21}^{+0.29}$ & 0.718 &   $3.99_{-0.41}^{+0.44}$ &  $0.60_{-0.18}^{+0.24}$    \\
MS0015.9+1609     &  $9.48_{-0.42}^{+0.57}$ & $0.19\pm0.05$ & 1.184 &   $9.98_{-0.48}^{+0.51}$ &  $0.23\pm0.05$  \\
WARPJ0848.8+4456  &  $1.27_{-0.37}^{+0.44}$ & $0.11_{-0.11}^{+0.23}$ & 0.407 &   $1.28_{-0.26}^{+0.29}$ &  $0.08_{-0.08}^{+0.13}$  \\
CLJ1354-0221      &  $2.18_{-0.68}^{+0.98}$ & $<0.40$ & 0.542 &   $2.71_{-0.94}^{+0.83}$  & $0.47_{-0.47}^{+0.75}$  \\
WARPJ1419.9+0634E &  $2.86_{-0.34}^{+0.37}$ & $0.29_{-0.18}^{+0.23}$ & 0.624 &   $2.93_{-0.23}^{+0.36}$ &  $0.32_{-0.16}^{+0.19}$  \\
MS0451.6-0305     &  $9.10_{-0.53}^{+0.56}$ & $0.14_{-0.06}^{+0.07}$ & 1.153 &   $9.26_{-0.39}^{+0.46}$ &  $0.18_{-0.05}^{+0.06}$  \\
RXJ0018.3+1618    &  $4.65_{-0.69}^{+0.52}$ & $0.31_{-0.16}^{+0.23}$ & 0.807 &   $3.87_{-0.41}^{+0.60}$ &  $0.47_{-0.22}^{+0.27}$   \\
WARPJ1419.3+0638  &  $3.48_{-0.41}^{+0.67}$ & $<0.15$ & 0.683 &   $3.93_{-0.47}^{+0.50}$  & $0.03_{-0.03}^{+0.13}$   \\
MS2053.7-0449     &  $3.94_{-0.50}^{+0.82}$ & $0.70_{-0.33}^{+0.55}$ & 0.725 &   $4.44_{-0.58}^{+0.69}$ &  $0.57_{-0.24}^{+0.35}$  \\
MACSJ0647.7+7015  &  $9.30_{-0.37}^{+0.45}$ & $0.27\pm0.05$ & 1.138 &   $10.52\pm0.40$ &  $0.37\pm0.04$  \\
CLJ1120+4318      &  $5.26_{-0.47}^{+0.49}$ & $0.34_{-0.12}^{+0.14}$ & 0.837 &   $5.22_{-0.37}^{+0.38}$ &  $0.52_{-0.12}^{+0.14}$  \\
CLJ1334+5031      &  $4.19_{-0.42}^{+0.44}$ & $0.16_{-0.12}^{+0.13}$ & 0.733 &   $4.73_{-0.41}^{+0.45}$ &  $0.15_{-0.09}^{+0.11}$  \\
MACSJ0744.9+3927  &  $8.14_{-0.34}^{+0.34}$ & $0.29\pm0.04$ & 0.995 &   $8.02_{-0.23}^{+0.24}$ &  $0.32\pm0.03$  \\
WARPJ1342.8+4028  &  $3.68_{-0.52}^{+0.57}$ & $0.68_{-0.28}^{+0.47}$ & 0.652 &   $3.60\pm0.47$ &  $0.58_{-0.23}^{+0.36}$  \\
CLJ1103.6+3555    &  $4.11_{-0.91}^{+1.06}$ & $<0.40$ & 0.659 &   $4.23_{-0.71}^{+0.95}$ &  $0.04_{-0.04}^{+0.18}$  \\
MS1137.5+6625     &  $6.12_{-0.97}^{+1.20}$ & $0.19_{-0.12}^{+0.16}$ & 0.813 &   $6.60_{-0.93}^{+0.92}$  & $0.37_{-0.17}^{+0.19}$  \\
CLJ1216.8-1201    &  $3.84_{-0.51}^{+0.75}$ & $0.26_{-0.19}^{+0.26}$ & 0.631 &  $4.40_{-0.52}^{+0.77}$ &  $0.09_{-0.09}^{+0.16}$   \\
MS1054.4-0321     &  $8.92_{-0.80}^{+0.85}$ & $0.15_{-0.10}^{+0.11}$ & 0.969 &  $8.85_{-0.74}^{+0.70}$ &  $0.17_{-0.09}^{+0.10}$   \\
CLJ0152.7-1357    &  $6.70_{-1.02}^{+1.19}$ & $0.23_{-0.16}^{+0.19}$ & 0.829 &  $6.58_{-0.78}^{+0.80}$ &  $0.24_{-0.13}^{+0.15}$    \\
CLJ1226.9+3332    & $10.16_{-0.73}^{+0.77}$ & $0.21\pm0.08$ & 0.998 &  $11.15\pm0.58$ &  $0.21_{-0.06}^{+0.07}$   \\
CLJ1429.0+4241    &  $4.10_{-0.77}^{+0.83}$ & $<0.12$ & 0.606 &  $4.75_{-0.51}^{+0.70}$ &  $<0.10$   \\
XLSSC029          &  $3.40_{-0.40}^{+0.53}$ & $0.38_{-0.18}^{+0.26}$ & 0.509 &  $4.06_{-0.42}^{+0.55}$ &  $0.56_{-0.20}^{+0.24}$   \\
RDCSJ1252-2927    &  $4.64_{-0.60}^{+0.66}$ & $0.24_{-0.14}^{+0.18}$ & 0.540 &  $4.84_{-0.53}^{+0.61}$ &  $0.33_{-0.14}^{+0.17}$   \\
1WGAJ2235.3-2557  &  $5.16_{-0.82}^{+1.00}$ & $0.14_{-0.14}^{+0.19}$ & 0.524 &  $5.87_{-0.76}^{+1.03}$ &  $0.12_{-0.12}^{+0.16}$   \\
\hline
\end{tabular}
\end{table*}

\vspace*{-0.5cm}
\subsection{Determination of $r_{500}$}

As an implication to the $M-T$ relation, the overdensity radius $r_{500}$ scales with the temperature of the cluster.
To compute $r_{500}$, we adopted the formula derived by \citet{vikhlinin06}
\begin{equation}
r_{500}hE(z)=0.792\left(\frac{\langle T \rangle}{5\: keV}\right)^{0.53}\: Mpc,
\end{equation}
where $E(z)=(\Omega_m(1+z)^3+\Omega_\Lambda)^{0.5}$.

\begin{figure}
\centering
\includegraphics[width=9.0 cm, angle=0]{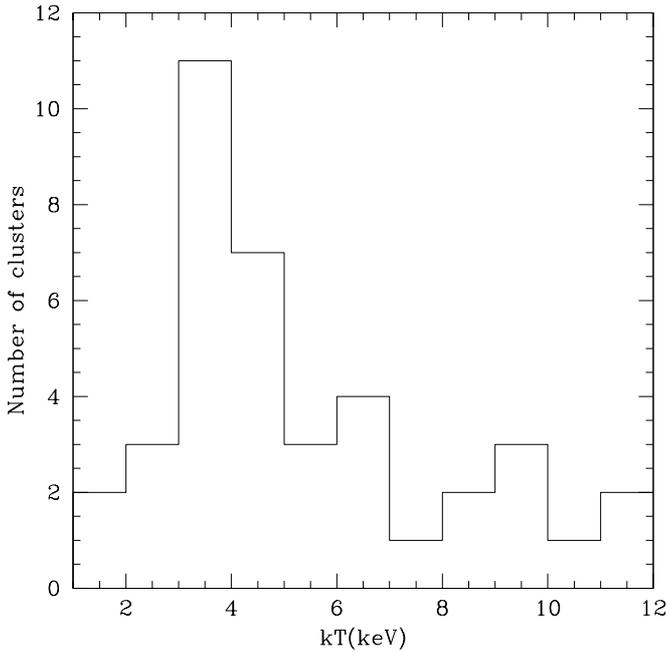}
\caption{Average temperature distribution (measured iteratively in the 0.15-0.6$r_{500}$ range) for
the clusters in our XMM-Newton sample.}
\label{histokT}
\end{figure}

To compute the global temperature
$\langle T \rangle$ necessary to estimate $r_{500}$, we extracted spectra
including emission going from 0.15$r_{500}$ to 0.5$r_{500}$ in each
cluster.  The central regions of each cluster are therefore excluded
from the spectra in order to avoid the contamination of a possible cool
core. The values of $\langle T \rangle$ and $r_{500}$ were
evaluated iteratively until a convergence to a stable value of the
temperature was obtained ($\Delta T\leq0.01$ keV between two successive
iterations).
Figure~\ref{histokT} shows the distribution of $<kT>$ in our sample, ranging from
$\sim1.5$~keV to $\sim11$~keV and peaking around $\sim4$~keV.

From the fits, we were also able to determine a global metallicity $\langle Z \rangle$ in each cluster.
In Table~\ref{xrayprop}, we list the best-fit values of $\langle T \rangle$ and $\langle Z \rangle$, 
and the value of $r_{500}$ computed using
the formula above. For completeness, in Table~\ref{xrayprop} we also report the values of the 
temperatures and abundance measured including the core emission. 

\subsection{Background treatment}
\begin{figure}
\centering
\includegraphics[width=9.0 cm, angle=0]{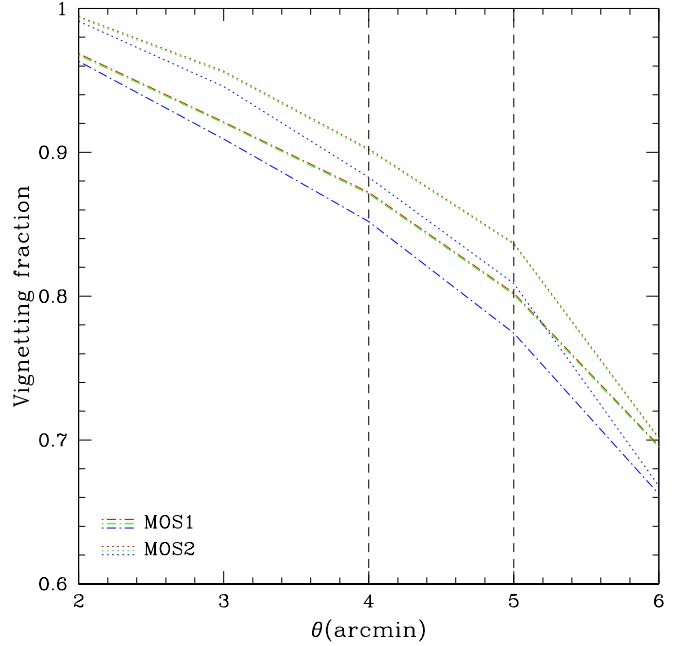}
\caption{Vignetting expected at different off-axis angle for the two MOS detectors. The red, green, and blue
lines represents the 0.7-1 keV, the 1-2 keV, and the 2-8 keV bands, respectively. The dashed black lines represents
the typical off-axis angle where we would extract the background for our sample. It is clear how the background spectra
could be vignetted up to a 20\% factor.}
\label{vign}
\end{figure}

The background of our MOS spectra was modeled (instead of subtracted) following the procedure developed
by \citet{leccardi08a}. These procedure is more accurate then a direct background subtraction because it 
allows us to use Cash statistics, which is recommended especially for low S/N spectra similar to those in
most of our sample \citep{nousek89}, and it enables us to avoid problems caused by the vignetting of the background spectra, which 
need to be extracted at larger off-axis angle than the cluster spectra ($\theta\sim5^\prime$) where the vignetting
of XMM mirrors is not negligible (Fig~\ref{vign}). Moreover, we note that this approach represents the most reliable 
characterization of the background (and its errors) available with respect to any other existing method.
The main aspects of this procedure are described below, and further details can be 
found on their paper.

We first extracted a spectrum in an annulus located between $10^\prime$ and 12$^\prime$ from the center of the field 
of view for both MOS1 and MOS2. To estimate the background parameters in this region (free of cluster emission), 
the resulting spectrum was fitted in XSPEC (in the 0.7-8 keV band, applying Cash statistics) using a model 
considering thermal emission from the Galaxy halo (HALO, XSPEC model: apec), the cosmic X-ray background (CXB, 
{\em XSPEC model: pegpwrlw}), a residual from the filtering of quiescent soft protons (QSP, {\em XSPEC model: bknpower}), 
the cosmic-ray-induced continuum ({\em NXB, XSPEC model: bknpower}), and the fluorescence emission lines ({\em XSPEC model: 
gaussian}). The response matrix file (RMF) of the detector was convolved with all the background components, while the 
ancillary response file (ARF) was convolved 
only with the first two components (HALO and CXB). The normalization of the QSP component was fixed at the value 
determined by measuring the surface brightness $SB_{in}$ in the 10$^\prime$-12$^\prime$ annulus, and comparing it 
to the surface 
brightness $SB_{out}$ calculated outside the field of view in the 6-12 keV energy band. Since soft protons are 
channeled by the telescope mirrors to within the field of view and the cosmic-ray-induced background covers the 
whole detector, the ratio $R_{SB}=\frac{SB_{in}}{SB_{out}}$ is a good indicator of the intensity of residual soft 
protons and was used in the modeling ($Norm(QSP) = 0.03 (1-R_{SB})$).
We also determined the 1$\sigma$ error for the background parameters to be used in fitting the cluster spectra.
To fit the cluster spectra, the background parameters (and their 1$\sigma$ errors) were then rescaled by the area 
from which the cluster spectra were extracted. Appropriate correction factors 
\citep[][dependent from the off-axis angle]{leccardi08a} were considered for the HALO and CXB components 
and a vignetting factor 
(corresponding to $1.858-0.078r$, where $r$ is the distance from the center of the annulus) was applied to the QSP 
component. All these rescaled values (and their 1$\sigma$ errors) were put in a XSPEC model that had the same background 
components as those used 
in the fit of the 10$^\prime$-12$^\prime$ annulus plus a thermal mekal model for the emission of the cluster whose 
temperature, abundance and normalization were allowed to vary. The 1$\sigma$ errors in the parameters 
were used to fix a range where the normalizations of the background components were allowed to vary.
This model was used to fit all the cluster spectra in our sample. Since we used Cash statistics, the 
spectra were minimally grouped to avoid spectral 
bins with zero counts. Moreover, a joint MOS1 plus MOS2 fit was used to increase the statistics. We were able to do that since
there are no calibration problems between the two detectors and an independent analysis of the spectra from MOS1 and MOS2
led to consistent results in the measures of both $kT$ and $Z$.

\begin{table*}
\caption{Comparison of fit results obtained with \citet{anders89} (ANGR89) and
\citet{asplund09} (ASPL09) solar abundance standards 
for four of the brightest clusters in our sample.\label{asplund}}
\centering
\begin{tabular}{|l|c|rrr|rrr|}
\hline
Cluster & Radial bin  &           &     ANGR89   &        &           &    ASPL09    &        \\
        & ($r_{500}$) & $kT$(keV) & $Z(Z_\odot)$ & C-stat & $kT$(keV) & $Z(Z_\odot)$ & C-stat \\
\hline\hline
RXJ2228.6+2037  & 0-0.15   &  $8.57_{-0.54}^{+0.51}$ & $0.27_{-0.07}^{+0.08}$ &  701.9 &  $8.58_{-0.54}^{+0.51}$ & $0.37_{-0.10}^{+0.12}$ &  702.4 \\
                & 0.15-0.4 &  $7.88_{-0.47}^{+0.52}$ & $0.27_{-0.06}^{+0.08}$ &  899.4 &  $7.89_{-0.47}^{+0.52}$ & $0.38_{-0.11}^{+0.13}$ &  899.2 \\
                & 0.4-0.9  &  $5.74_{-0.40}^{+0.44}$ & $0.12\pm0.07$          & 1032.4 &  $5.72_{-0.39}^{+0.43}$ & $0.17_{-0.10}^{+0.11}$ & 1032.4 \\
\hline
RXCJ1206.2-0848 & 0-0.15   &  $9.91_{-0.46}^{+0.50}$ & $0.28_{-0.05}^{+0.06}$ &  878.2 &  $9.95_{-0.47}^{+0.51}$ & $0.39_{-0.07}^{+0.08}$ &  878.5 \\
                & 0.15-0.4 & $11.61_{-0.65}^{+0.81}$ & $0.18\pm0.07$          & 1120.6 & $11.61_{-0.63}^{+0.72}$ & $0.26_{-0.09}^{+0.10}$ & 1120.6 \\
                & 0.4-0.7  &  $9.20_{-0.83}^{+1.26}$ & $0.50_{-0.14}^{+0.15}$ & 1109.7 &  $9.15_{-0.82}^{+1.19}$ & $0.72\pm0.21$          & 1109.1 \\
\hline
RXJ1347.5-1145  & 0-0.15   & $11.11\pm0.26$          & $0.34_{-0.04}^{+0.03}$ & 1204.2 & $11.14\pm0.26$          & $0.48\pm0.05$          & 1205.0 \\
                & 0.15-0.4 & $11.55_{-0.47}^{+0.48}$ & $0.25_{-0.05}^{+0.06}$ & 1072.5 & $11.57_{-0.46}^{+0.48}$ & $0.35_{-0.07}^{+0.08}$ & 1073.1 \\
                & 0.4-0.8  &  $7.74_{-0.64}^{+0.84}$ & $0.15_{-0.08}^{+0.09}$ & 1075.3 &  $7.57_{-0.59}^{+0.82}$ & $0.21_{-0.11}^{+0.13}$ & 1076.6 \\
\hline
MS0451.6-0305   & 0-0.15   &  $9.47_{-0.57}^{+0.81}$ & $0.24_{-0.08}^{+0.09}$ &  631.3 &  $9.49_{-0.57}^{+0.81}$ & $0.34_{-0.12}^{+0.13}$ &  631.2 \\
                & 0.15-0.4 &  $9.47_{-0.56}^{+0.67}$ & $0.18\pm0.07$          &  786.8 &  $9.46_{-0.56}^{+0.68}$ & $0.25_{-0.10}^{+0.11}$ &  786.7 \\
                & 0.4-0.7  &  $7.29_{-0.86}^{+1.04}$ & $0.07_{-0.07}^{+0.11}$ &  806.2 &  $7.25_{-0.85}^{+1.02}$ & $0.10_{-0.10}^{+0.17}$ &  806.3 \\
\hline
\end{tabular}
\end{table*}

\subsection{Effects of the XMM-Newton PSF on the analysis}\label{psf}
In contrast with Chandra, the EPIC XMM-Newton telescopes have a quite large PSF ($HEW\sim15^{\prime\prime}$).
This was a potential problem to our analysis because the emission from the cluster cores 
($r<0.15r_{500}$) in principle could contaminate the spectra just outside the core ($0.15r_{500}<r<0.4r_{500}$)
and viceversa,
if the extraction region for the cores is similar in size to the HEW or smaller, as happens in about
a third of our sample and in almost all the clusters at $z>0.7$. However, the contribution of 
the emission from the $0.15-0.4r_{500}$ annulus to the outer region spectra ($r>0.4r_{500}$) can be assumed to be 
negligible. If the source is uniform over the field, then this effect is not a major concern.
However, if there are strong gradients in the emission and spectral parameters over the field (as it is the 
case of cooling core clusters) then the effect might be significant.

To quantify the effect of the PSF on the determination of the spectral parameters measured (and especially
on the abundance measurements), we used a procedure developed in XMM-ESAS \citep{snowden04} that considers 
the crosstalk ARFs between two contiguous regions to remove the contribution of the core from the surrounding
region and viceversa. We applied it to a few clusters in the sample
(RXJ1213.5, CLJ1354, MS1054.4, and CLJ1429.0) at different redshifts and core extraction radii, which shows evidence
of a temperature dip at the center and abundance gradients.
In all of these clusters, the differences in the temperatures measured considering the PSF correction and the
temperatures measured when not considering the correction were in the range between 5\% and 15\%, which are 
comparable to or smaller than
the statistical errors in the measurements. On the other hand, the differences in the abundance measures were always $\leq0.02$
$Z_\odot$, i.e. well below the statistical errors.
Therefore, we decided not to apply the correction to the remainder of the sample.
\begin{figure*}
\centering
\includegraphics[width=9.0 cm, angle=0]{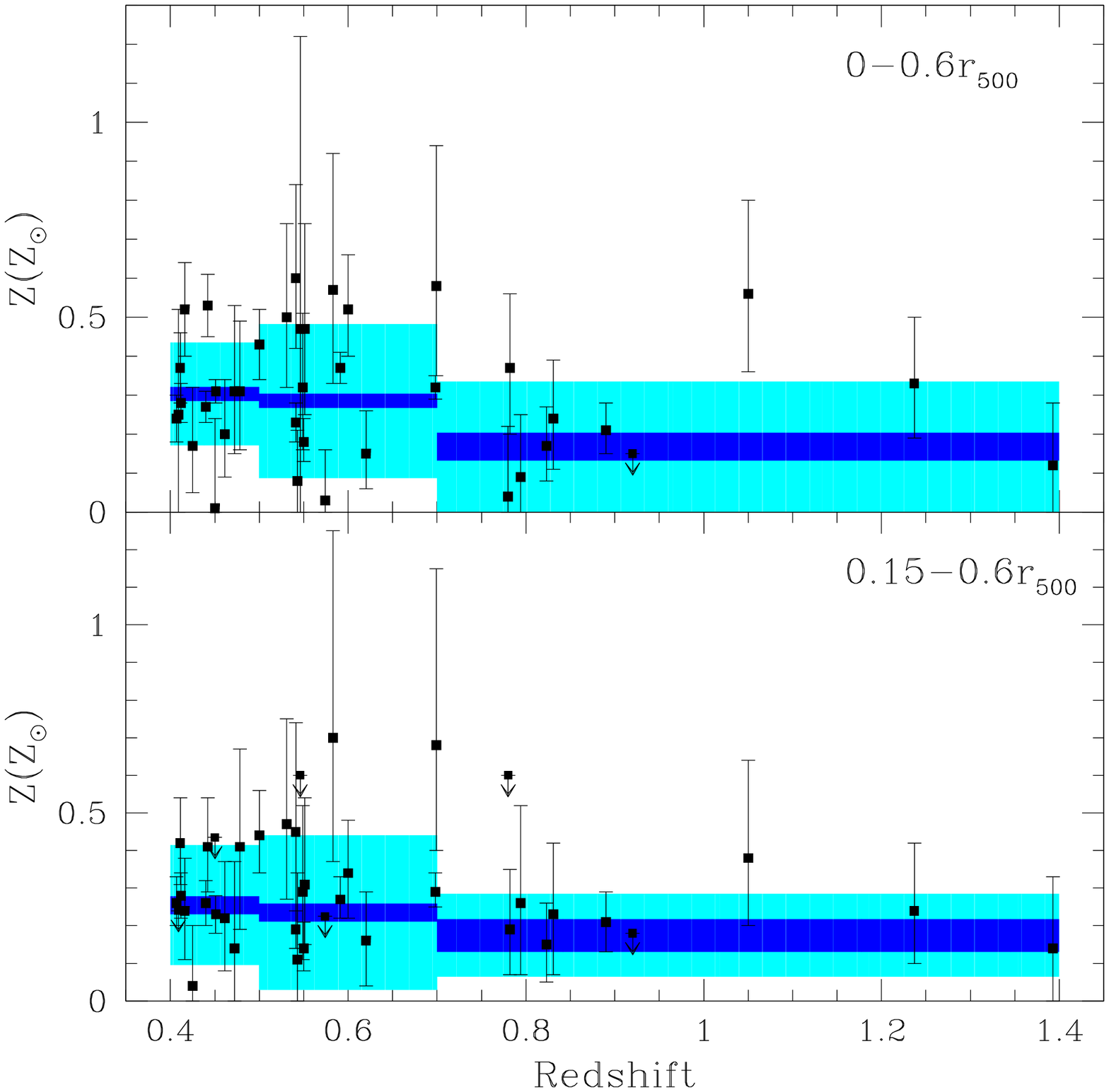}
\includegraphics[width=9.0 cm, angle=0]{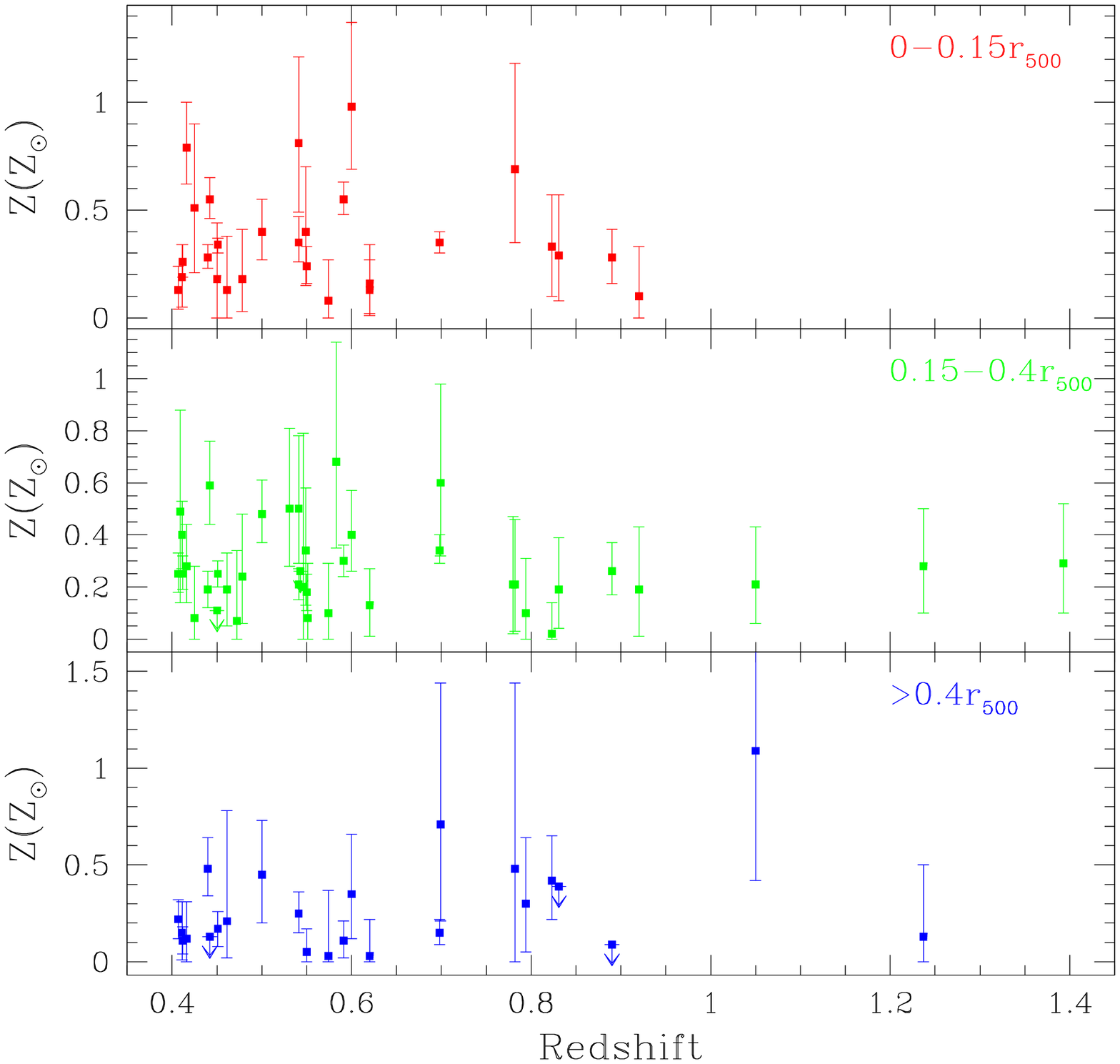}
\caption{{\it Left:} Abundance measured by EPIC XMM-Newton as a function of the redshift for the
clusters in our sample including the emission from the whole cluster at $r<0.6r_{500}$ (top) and excising
the core (bottom). The error bars in $Z$ are at 1$\sigma$, while
shaded areas show the weighted mean of the abundance with its error (blue) and rms dispersion (cyan) in three redshift bins.
{\it Right:} Abundance measured by EPIC XMM-Newton as a function of the redshift for the
clusters in our sample in different spatial bins (from top to bottom: 0-0.15$r_{500}$,0.15-0.4$r_{500}$, and $>0.4r_{500}$),
with 1$\sigma$ error bars indicated.}
\label{zabund}
\end{figure*}

\subsection{Spectral fitting}

The spectra were analyzed with XSPEC v12.5.1 \citep{arnaud96} and fitted
with a single-temperature {\tt mekal} model \citep{liedahl95} with
Galactic absorption ({\tt wabs} model), in
which the ratio between the elements was fixed to the solar value as in
\citet{anders89}. We prefer to
report metal abundances in these units of solar abundances
because most of the literature still refers to them.
We tested the robustness of our results to variations in the assumed emission 
abundance model by fitting the data for four of the brightest clusters in our
sample (RXJ2228.6+2037, RXCJ1206.2-0848,
RXJ1347.5-1145, and MS0451.6-0305) using also the most recent values of \citet{asplund09}.
Table~\ref{asplund} shows the results of the fits obtained using the two
different solar abundance references.
The fits were of similar quality, as can be seen from the similar values of the C-statistic, 
and the only significant differences were those found in the values of the fitted abundances, which we found
to be consistent with a simple scaling by a factor of 1.41 (within a 3\% margin), 
where the \citet{asplund09} abundances are higher than those of \citet{anders89}. 
This is a clear indication that we have measured the iron abundance in the ICM, as
expected from the statistics we have in virtually all the cluster spectra.
In addition, also the fitted temperatures using the two abundance models were all consistent.


All the spectral fits were performed over the energy range $0.7-8.0$~keV for both 
of the two MOS detectors. We fitted simultaneously MOS1 and MOS2 spectra
to increase the quality of the statistics.


The free parameters in our spectral fits are temperature, abundance,
and normalization. Local absorption is fixed to the Galactic neutral
hydrogen column density, as obtained from radio data  \citep{kalberla05}, 
and the redshift to the value measured from optical spectroscopy ($z$ in 
Table~\ref{exposures}). We used Cash statistics applied to the source plus
background\footnote{http://heasarc.gsfc.nasa.gov/docs/xanadu/xspec/manual/
XSappendixStatistics.html}, 
which is preferable for low S/N spectra \citep{nousek89}
and requires a minimal grouping of the spectra (at least one count per spectral bin).
We estimated a goodness-of-fit for the best-fit models using the 
XSPEC command {\tt goodness}, which simulates a user-defined number of spectra based 
on the best-fit model itself and calculates the percentage of these simulations 
in which the fit statistic is smaller than that for the data.
For each spectrum in the sample, we simulated 10000 spectra from the best-fit model, 
whose C-statistic values were compared with the value obtained from the original spectrum.
For the spatially resolved bins ($r=0-0.15r_{500}$, $r=0.15-0.4r_{500}$ and
$r>0.4r_{500}$), we found that the percentage of simulations with
a smaller value of the C-statistic than that for the data is always less than 40\%, 
an indication that the one-temperature model 
is always a good fit to the spatially resolved spectra.
For the spectra extracted including emission from the whole cluster ($r=0-0.6r500$  
and $r=0.15-0.6r500$), the percentage in a few cases is found to be larger than 40\%, 
especially in the spectra with more counts. 
This is fully expected because the potential well of galaxy clusters is very likely 
to harbor a multi-temperature gas. In those cases, the one-temperature model 
is not a good fit to the data, but this is not a major problem because we simply wish to measure 
an average temperature and abundance for the integrated emission.

\section{Results}
The main results we attained from the spectral analysis
are presented in this Section. In particular, we focus on the redshift evolution of the metal abundance
considering both the integrated and spatially resolved emission from the clusters, and on the correlation
of the temperature with both redshift and abundance.

\subsection{Integrated abundance evolution}\label{abundev}

Although the main focus of this paper has been to spatially resolve the evolution of the abundance with redshift, a complementary
measurement is an integrated measure of the abundance, which is useful when comparing with previous works in the 
literature.

Figure~\ref{zabund} (left) shows how the measured abundance in our cluster sample evolves 
with redshift. To perform a statistical verification of this evolution, we computed a Spearman's rank $\rho$ for the distributions
plotted in the graph considering both emission from the whole cluster ($r=0-0.6r_{500}$) and after excising the cores
($r=0.15-0.6r_{500}$). We performed 10000 Monte Carlo (MC) simulations 
to take into account the distribution of the 
errors in $Z$ in the determination of the Spearman's rank $\rho$. 
In the distribution considering emission from the whole cluster, we observed a weak negative correlation between redshift and
abundance with
a mean Spearman's rank $\rho=-0.12\pm0.10$ with 39 degrees of freedom (d.o.f.), corresponding to a mean probability of no
correlation $p=0.48\pm0.32$. The negative correlation was found to be even weaker after we had excised the core, as could be deduced 
from the computed mean Spearman's rank $\rho=-0.05\pm0.11$ (again with 39 d.o.f.), which corresponds to a probability of no correlation
$p=0.61\pm0.31$. 
We also fitted the data points in Figure~\ref{zabund} (left) with a power-law such as $Z\propto (1+z)^{-\gamma}$, where $Z$
is the abundance measured and $z$ is the cluster redshift.
The value of $\gamma$ (reported in Table~\ref{compgamma})
indicates a weak negative evolution with a significance lower than 2$\sigma$ in both cases. The most significant
deviation from $\gamma=0$ is observed in the $0-0.6r_{500}$ sample, where $\gamma=0.75_{-0.74}^{+0.80}$ at a 90\% confidence
level. In the sample where the core is excised, the power-law slope is consistent with zero ($\gamma=0.61_{-0.66}^{+0.72}$).

\subsection{Spatially resolved abundance evolution}\label{abundres}

As stated before, the main goal of our paper has been to study the evolution of the metal abundance in galaxy clusters through 
cosmic time, trying to spatially resolve the cluster emission into three radial bins. 
Figure~\ref{zabund} (right) shows how the abundance measured at different radii in the cluster sample evolves 
with redshift.

Similarly to the steps taken in \S~\ref{abundev}, we used two different statistical methods to quantify any presence of 
evolution: the calculation of a mean Spearman's rank $\rho$ (averaged performing 10000 MC simulations) and a power-law 
fit to the data points.

In the $r<0.15r_{500}$ radial bin, we do not observe any evidence of abundance evolution
with a mean Spearman's rank $\rho=0.03\pm0.14$ with 26 degrees of freedom (d.o.f.), corresponding to a mean probability of no
correlation $p=0.62\pm0.28$. 
Fitting the data points in this bin with a power-law in the form $Z\propto (1+z)^{-\gamma}$, as in \S~\ref{abundev}, we obtained
$\gamma=-0.19_{-0.68}^{+0.72}$ (Table~\ref{compgamma}), indicating no significant evolution in agreement with the mean Spearman's rank
result.

In addition, at $0.15r_{500}<r<0.4r_{500}$, the values are very widely dispersed as is clear
from the computed mean Spearman's rank $\rho=-0.06\pm0.12$ with 39 d.o.f., which corresponds to a probability of no correlation
$p=0.54\pm0.31$. In this case, the value of $\gamma$ from the power-law fit is positive ($\gamma=0.44_{-0.72}^{+0.78}$, as shown in 
Table~\ref{compgamma}), although it is
consistent with zero within the 1$\sigma$ errors, confirming that no significant abundance evolution was found in this radial bin.

The correlation is not present at larger radii ($r>0.4r_{500}$) with a mean Spearman's rank $\rho=0.08\pm0.16$ and a 
mean $p=0.54\pm0.33$. Fitting the data points with a power-law, the value of $\gamma$ is positive but also consistent with zero
within $1\sigma$ in this case ($\gamma=1.65_{-1.92}^{+2.33}$, as reported in Table~\ref{compgamma}).

\begin{figure}
\centering
\includegraphics[width=9.0 cm, angle=0]{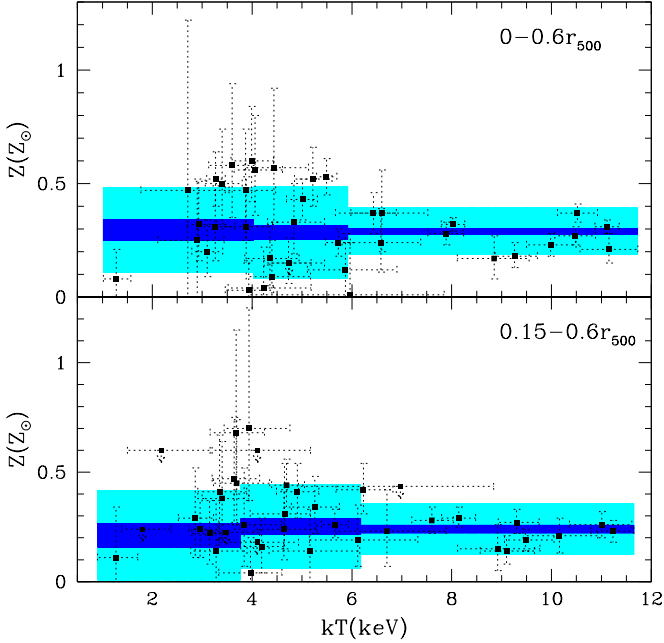}
\caption{Temperature vs. Abundance measured for the galaxy clusters in our sample including (top) or
excluding the core (bottom). The dotted lines represent the 1$\sigma$ errors in $kT$ and $Z$, while
shaded areas show the weighted mean in the abundance with its error (blue) and rms dispersion (cyan) in three temperature bins.
Althought the values of $Z$ show a larger spread at $kT<5$ keV, we do not find any correlation between the two quantities.}
\label{kTabund}
\end{figure}

\subsection{Temperature-abundance and temperature-redshift correlation}
As stated in \S~\ref{psf}, the PSF of EPIC could in principle have an effect on the determination of the
temperature in the central two annuli of the clusters (especially at high redshift) 
of the order of 5-15\%, if there were temperature gradients. Therefore we decided to compare the distribution 
of $Z$ as a function of the temperature only for the whole cluster emission (or after simply excising the core as 
in \S~\ref{abundev}).

Figure~\ref{kTabund} shows the correlation between the temperature and the abundance measured in each
cluster including ($r=0-0.6r_{500}$) and excluding the core ($r=0.15-0.6r_{500}$). Although at first glance, 
it appears that higher values of abundance
are more frequent at low temperatures, the correlation 
between this two quantities is statistically very weak (as can be clearly seen from the weighted average values of $Z$
derived in three different bins of temperature).0
When we considered the emission from the whole cluster, we measured a Spearman's rank 
$\rho=-0.09\pm0.13$ with 39 d.o.f. (errors computed using 10000 MC simulations as in \S~\ref{abundev}), corresponding to
a probability of no correlation $p=0.52\pm0.33$. The correlation was also found to be weak after we had excised the core
($\rho=0.02\pm0.13$ with 39 d.o.f., $p=0.56\pm0.30$).
We also fit the distribution with a power-law of the form $Z\propto kT^{-\gamma}$. The results are consistent with
there being no correlation between $kT$ and $Z$ for both samples ($\gamma=-0.06\pm0.11$ including the core, and $\gamma=-0.06\pm0.16$
excising the core).
Since, as stated before, we found that the values of $Z$ have a greater dispersion at lower temperatures, we divided
each sample into two subsamples at $kT<5$ keV and $kT>5$ keV and compared them with a Kolmogorov-Smirnov (K-S) test to verify whether
they come from the same distribution. Both including and excising the core, the K-S test cannot reject at a 5\% confidence level
the null-hypothesis that the clusters with $kT<5$ keV have the same distribution of $Z$ as the clusters with $kT>5$ keV.
We therefore do not find any obvious trend in the abundance with temperature similar to those seen in other samples of galaxy clusters 
observed with other X-ray missions such as Chandra (BLS07) and ASCA \citep[e.g.][]{baumgartner05}. 

\begin{figure}
\centering
\includegraphics[width=9.0 cm, angle=0]{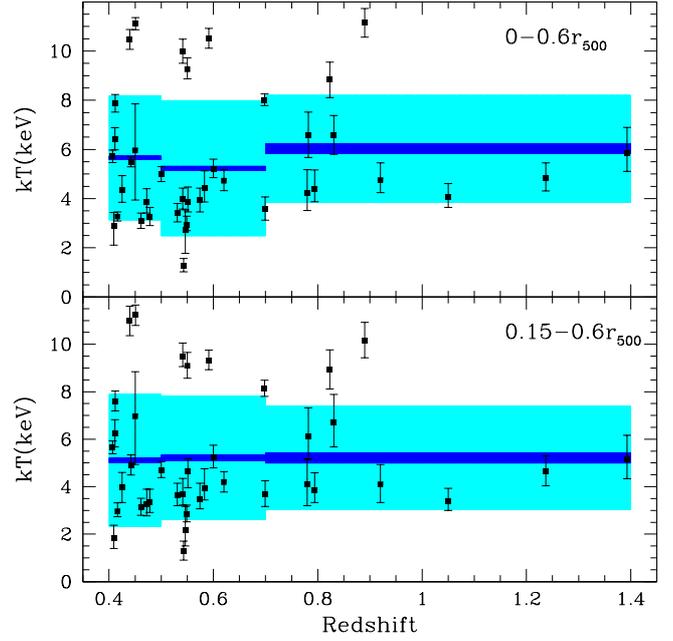}
\caption{Redshift vs. temperature measured for the galaxy clusters in our sample including (top) or
excluding the core (bottom). The errors in $kT$ are plotted at 1$\sigma$, while
shaded areas show the weighted mean of the temperature with its error (blue) and rms dispersion (cyan) in three redshift bins.
No clear correlation between redshift and temperature is observed.}
\label{zkT}
\end{figure}

No clear correlation is also observed in the redshift versus (vs.) temperature distribution (Figure~\ref{zkT}),
as indicated by a Spearman's rank $\rho=0.14\pm0.11$ with 39 d.o.f., (probability of no correlation $p=0.39\pm0.25$), for
the $0-0.6r_{500}$ sample, and a Spearman's rank $\rho=0.09\pm0.11$ (39 d.o.f., $p=0.51\pm0.26$) for
the $0.15-0.6r_{500}$ sample. Figure~\ref{zkT} also shows that we have analyzed a population of clusters whose temperatures
are uniformly distributed with redshift, there being no preferential redshift range containing the hottest clusters in 
the sample ($kT>8$~keV).

\begin{table*}
\caption{Values of $\gamma$ in the power-law fits of the $z$-$Z$ distribution ($Z\propto (1+z)^{-\gamma}$) in the samples
present in the literature compared with those obtained in the present work. The errors in $\gamma$ and the normalization
(referred at $z=0.6$) are quoted at 1$\sigma$ for one interesting parameter.
\label{compgamma}}
\centering
\begin{tabular}{|l|c|c|c|c|c|}
\hline\hline
 & Extraction radius & $N_{obj}$ & Redshift range & $\gamma$ & Norm (@$z=0.6$)\\
\hline
This work (\S~\ref{abundev})  & $0-0.6r_{500}$    & 39 & 0.407--1.393 & $0.75_{-0.45}^{+0.48}$ & $0.28\pm0.01$ \\
                              & $0.15-0.6r_{500}$ & 39 & 0.407--1.393 & $0.61_{-0.66}^{+0.72}$ & $0.24\pm0.02$ \\
\hline
This work (\S~\ref{abundres}) & $0-0.15r_{500}$   & 26 & 0.407--0.920 & $-0.19_{-0.68}^{+0.72}$ & $0.33\pm0.02$ \\
                              & $0.15-0.4r_{500}$ & 39 & 0.407--1.393 & $0.44_{-0.72}^{+0.78}$ & $0.23\pm0.02$ \\
                              & $>0.4r_{500}$     & 24 & 0.407--1.237 & $1.65_{-1.92}^{+2.33}$ & $0.13\pm0.03$ \\
\hline
BLS07                         & $0.15-0.3r_{vir}$ & 46 & 0.405--1.273 & $0.68_{-0.47}^{+0.53}$ & $0.28\pm0.01$ \\
\hline
MAU08\tablefootmark{a}        & $0-1r_{500}$      & 50 & 0.405--1.237 & $2.44_{-0.69}^{+0.76}$ & $0.30\pm0.02$ \\
                              & $0.15-1r_{500}$   & 46 & 0.405--1.237 & $3.87_{-0.92}^{+1.06}$ & $0.22\pm0.02$ \\
\hline
MAU08\tablefootmark{b}        & $0-1r_{500}$      & 49 & 0.405--1.237 & $2.18_{-0.71}^{+0.77}$ & $0.31\pm0.02$ \\
                              & $0.15-1r_{500}$   & 45 & 0.405--1.237 & $-0.15_{-1.23}^{+1.39}$ & $0.28\pm0.02$ \\
\hline
\citet{anderson09}            & $0-X^{\prime\prime}$\tablefootmark{c} & 23 & 0.407--1.237 & $0.09_{-0.76}^{+0.88}$ & $0.24\pm0.02$ \\
\hline
\end{tabular}
\tablefoot{
\tablefoottext{a}{Including CLJ1415.1+3612 in the fit.}\\
\tablefoottext{b}{Not including CLJ1415.1+3612 in the fit.}\\
\tablefoottext{c}{Maximum extraction radius $X$ listed in their paper and determined using S/N optimization criteria.}
}
\end{table*}

\section{Discussion}
The main goal of our paper has been to study the evolution of the metal abundance in galaxy clusters through cosmic time. To achieve
this aim, we did not consider the emission only from the entire cluster (\S~\ref{abundev}) but we also tried to spatially resolve the 
cluster emission into three radial bins of $0-0.15r_{500}$, $0.15-0.4r_{500}$, and $>0.4r_{500}$ (\S~\ref{abundres}).

\subsection{Comparison with previous works}\label{compprev}

In \S~\ref{abundev}, we have shown that considering the integrated emission from the whole clusters we could detect only a mild 
negative abundance evolution with redshift at a confidence level lower than 2$\sigma$ 
(as indicated by the power-law fits of the $z$-$Z$ distribution).

\begin{figure}
\centering
\includegraphics[width=9.0 cm, angle=0]{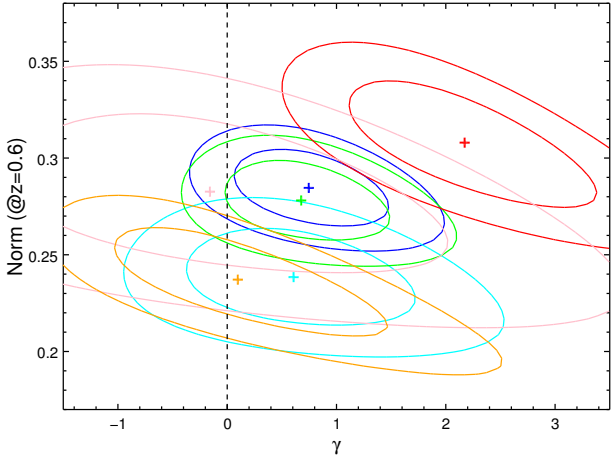}
\caption{Confidence contours (computed at 1$\sigma$ and 2$\sigma$ for two interesting parameters) 
on $\gamma$ and normalization (referred to z=0.6) for all the datasets analyzed in \S~\ref{compprev}.
Blue and cyan contours refer to our $0-0.6r_{500}$ sample and our $0.15-0.6r_{500}$ sample, respectively.
The contours for MAU08 samples (computed excluding CLJ1415.1+3612, see text and Table~\ref{compgamma}) at
$r=0-1r_{500}$ and $r=0.15-1r_{500}$ are plotted in red and pink, respectively. Green contours refer
to the BLS07 sample, while orange contours are plotted for the \citet{anderson09} dataset.
}
\label{ctrplots}
\end{figure}

Notwithstanding, it is interesting to perform a quantitative comparison of our results with those obtained previously in
the literature (e.g. BLS07).
The Chandra samples of BLS07 and MAU08 and the XMM-Newton sample of \citet{anderson09} are the
most suitable to perform this comparison. In these samples, we selected
only the clusters at $z>0.4$ to be consistent with the redshift range probed in our work. We have 23, 46, 50, and 46 objects
in this redshift range for the \citet{anderson09} sample, the BLS07 sample, the MAU08 sample, and the MAU08 sample with the
core excised ($r<0.15r_{500}$), respectively. We performed the same power-law fit to the $z$-$Z$ relation that we used in 
\S~\ref{abundev} for our data (in the form $Z\propto (1+z)^{-\gamma}$). The comparison of the fits with those obtained
for our sample are shown in Table~\ref{compgamma} and Figure~\ref{ctrplots}, where the confidence contours 
on $\gamma$ and normalization (referred at z=0.6) are shown for all the datasets analyzed. 
From the values of $\gamma$ and normalization shown in the table and the confidence contours shown in Figure~\ref{ctrplots},
it is clear how our results for the integrated measure of the abundance in the $0-0.6r_{500}$ sample (\S~\ref{abundev}) 
are fully consistent with BLS07 sample. Both these samples do not display any significant signs of negative evolution
(significance lower than 2$\sigma$). Table~\ref{compgamma} also shows that the sample of \citet{anderson09} has a 
$\gamma\sim0$, while the only 
samples showing an evolution of abundance with $z$ at a confidence level higher than 2$\sigma$ are the MAU08 Chandra samples. 
In particular, it is striking how the values of $\gamma$ from the MAU08 sample are significantly higher
than in all the other samples (especially not considering the core emission). We note that these values are probably affected by 
the presence in their sample of a single cluster at high redshift with a very low upper limit measurement on $Z$
(CLJ1415.1+3612 at $z=1.030$ with $Z<0.14$ and $Z<0.04$, including and excising the core, respectively).
Therefore we performed the power-law fits of their sample excluding this particular cluster to quantify its influence on the
$z$ vs. $Z$ trends. While in the sample including the core, the results of the evolution strength are quite robust 
($\gamma$ changes from $2.44_{-0.69}^{+0.76}$ to $2.18_{-0.71}^{+0.77}$), removing CLJ1415.1+3612 from the $0.15-1r_{500}$ sample
has a dramatic effect on the measured abundance evolution. The value of $\gamma$ indeed changes from $3.87_{-0.92}^{+1.06}$, 
corresponding to a strong negative evolution, to $-0.15_{-1.23}^{+1.39}$, corresponding to no significant evolution. This is also 
evident from Figure~\ref{ctrplots}, where the confidence contours in the $\gamma$-normalization parameter space 
referring to this sample are consistent with no evolution and with all the other datasets considered in this section. 
That a single deviant object could have such a strong effect on the abundance evolution significance is clearly a caveat 
about being careful in interpreting results based on such small samples.
In any case, from the values reported in Table~\ref{compgamma} and the projection of the contours in Figure~\ref{ctrplots}, 
we can constrain the normalization (at $z=0.6$) to $Z\approx0.28$ and $Z\approx0.24$ in our $0-0.6r_{500}$ sample and in our 
$0.15-0.6r_{500}$ sample, respectively. These values are consistent within 1$\sigma$ with the values obtained from all 
the other datasets in the literature considered in the present section.
We note that the lack of evolution in almost all the samples analyzed in this section could be due, at least in part, 
to the redshift range probed in this paper. BLS07, MAU08 and \citet{anderson09} did indeed detect a significant 
evolution in abundance, only between the low redshift clusters ($0.1\la z\la 0.3$) and the clusters at $z\gtrsim 0.5$.

As a consistency check, we also applied a Bayesian MC method to fit the datasets described above \citep[see e.g.][]{andreon10}, finding
consistent results with the power-law minimum least squares fit for all the samples.


\begin{figure}
\centering
\includegraphics[width=9.0 cm, angle=0]{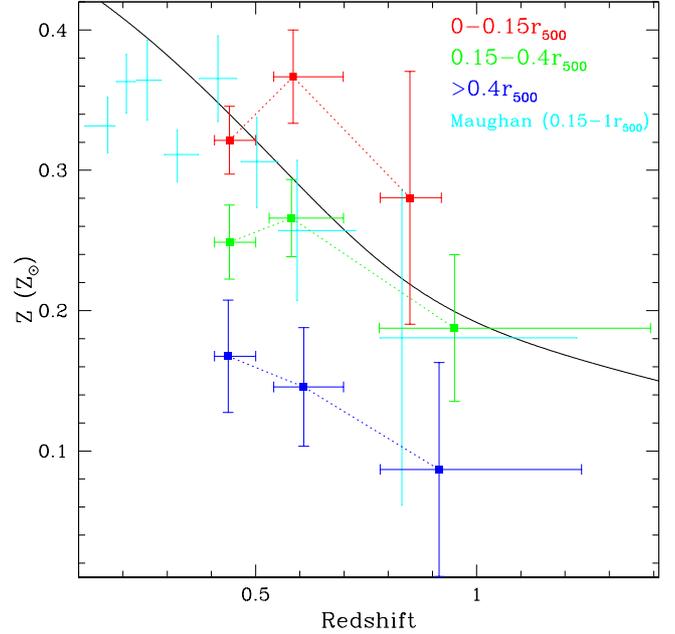}
\caption{Mean weighted abundance in our XMM-Newton sample as a function of the redshift for the
clusters in our sample in different spatial bins: 0-0.15$r_{500}$ plotted in red, 0.15-0.4$r_{500}$ in green, 
and $>0.4r_{500}$ in blue. The error bars represent the error in the mean and are listed in Table~\ref{medabund}. 
The cyan points represent the weighted abundance measured by MAU08 for a
sample of 116 clusters. The abundance values expected at different redshifts in the \citet{ettori05} model 
are plotted as a solid black line. 
}
\label{zabundmed}
\end{figure}

\begin{table*}
\caption{Mean weighted abundance in the three radial bins considered 
computed at three different redshift bins, with their errors and r.m.s. dispersions indicated.\label{medabund}}
\centering
\begin{tabular}{|l|c|c|c|}
\hline
Redshift & $\langle Z(0-0.15r_{500})\rangle \pm err \pm rms$  & $\langle Z(0.15-0.4r_{500})\rangle \pm err \pm rms$ & 
$\langle Z(>0.4r_{500})\rangle \pm err \pm rms$\\
\hline
0.4--0.5 & $0.32\pm0.02\pm0.19$ & $0.25\pm0.03\pm0.17$ & $0.17\pm0.04\pm0.16$ \\
0.5--0.7 & $0.37\pm0.03\pm0.29$ & $0.27\pm0.03\pm0.20$ & $0.15\pm0.04\pm0.23$ \\
0.7--1.4 & $0.28\pm0.09\pm0.20$ & $0.19\pm0.05\pm0.08$ & $0.09\pm0.08\pm0.44$ \\
\hline
\end{tabular}
\end{table*}
\subsection{Spatially resolved abundance: weighted mean and radial dependence of $Z$}

In \S~\ref{abundres}, we resolved spatially the emission from our clusters into three radial bins.
However, no evolution in abundance could be observed in each of the radial bins considered. In this case, the deviations from
no evolution are even less significant than in the integrated measures we have obtained in this paper (\S~\ref{abundev}) and in
the results of BLS07 and MAU08. The most likely reasons for this discrepancy 
are the larger errors in the measure of $Z$ and the smaller sample size of the $r=0-0.15r_{500}$ and the
$r>0.4r_{500}$ samples (26 and 24 clusters, respectively). Both of these reasons are clearly caused by the smaller number of counts
we have when dividing the emission into three radial bins instead of considering a single spatial bin for each cluster.

Although the low quality of our statistics prevent us from any quantitative claim about the abundance evolution, it is interesting to 
consider a weighted mean of the abundance computed at different redshift bins to perform a qualitative comparison with the literature 
(in particular with MAU08) and the theoretical predictions of the \citet{ettori05} model.
The size of our XMM-Newton sample did not allow us to consider more than three redshift bins 
($0.4<z<0.5$, $0.5<z<0.7$, and $z>0.7$) when computing the weighted mean and using as weights the statistical errors in 
the measure of the abundance.
Figure~\ref{zabundmed} shows the mean weighted abundance in the three radial bins considered 
computed at the three different redshift bins listed above. The mean weighted abundances are also shown in Table~\ref{medabund}
with their errors and r.m.s. dispersions. We have also plotted the weighted means of MAU08 for their
116 Chandra clusters (obtained by performing a stacking spectral analysis and excising the cores) 
and the predictions of the \citet{ettori05} model for the abundance evolution in galaxy clusters.
As expected, the abundance weighted-mean decreases towards larger radii (consistently with the power-law 
normalizations reported in Table~\ref{compgamma}).
However, the $z$-$Z$ relation is consistent with no evolution at every radius. As stated before, the small number of 
counts in each radial bin (and the associated large errors in $Z$) for most of the clusters in the sample does not allow us to 
measure with sufficient accuracy the mean abundance at different redshift, and prevents any statistically significant claim 
about its evolution with cosmic time in the three different cluster regions considered 
(i.e. the core, the core surroundings, and the outskirts), in agreement to what we found in \S~\ref{abundres}.
From Figure~\ref{zabundmed}, we can see that, although there is no supporting statistical evidence of any kind of evolution, 
the trend in the 0.15-0.4$r_{500}$ radial bin generally complements nicely the measures of MAU08,
obtained after excising the cool core ($0.15r_{500}<r<r_{500}$), and broadly agrees with the predictions of the \citet{ettori05} model.
In particular, the model of \citet{ettori05} predicts a slope between $z=0.4$ and $z=1.4$ of $\gamma=1.7$, which is only
$\sim1.5\sigma$ above the constraints we have in our 0.15-0.4$r_{500}$ dataset.

The dependence of the abundance $Z$ on both the radius $r$ and the redshift $z$ could also be tested by fitting 
$Z$, $r$, and $z$ from the 89 data points in the three radial bins together 
(the points plotted in the right panel of Figure~\ref{zabund}) with a function of two variables in the form
\begin{equation}
Z(r,z)=Z_0\left(1+\left(\frac{r}{r_0}\right)^2\right)^{-a}\left(\frac{1+z}{1.6}\right)^{-\gamma}.
\end{equation}
In this formula, the radius of reference $r$ in each radial bin was defined as the
effective radius $\bar{R}$ weighted by the emissivity distribution within a given radial bin defined by
\begin{equation}
S_b (R_i, R_o) = \frac{2}{(R_o^2-R_i^2)} \int_{R_i}^{R_o} S_b \, r \, dr = S_b(\bar{R}),
\end{equation}
where $S_b$ is the X-ray surface brightness, and
$R_i$ and $R_o$ are the inner and outer radius of the radial bin. Under the assumption that $S_b \propto r^{-\alpha}$, we obtain
\begin{equation}
\bar{R} = \left( \frac{2}{(2-\alpha)} \times \frac{(R_o^{2-\alpha} -R_i^{2-\alpha})}{(R_o^2 -R_i^2)} \right)^{-1/\alpha}.
\end{equation} 
In our estimates, we adopt $\alpha=$ 0.7, 2.2, and 3 in the $0-0.15$, $0.15-0.4$, and $>0.4 r_{500}$ radial bin, respectively, 
in accordance with the expected radial  behaviour of the X-ray surface brightness in massive galaxy clusters 
\citep[see e.g.][]{ettori04}. Fixing $r_0=0.15r_{500}$, we obtained a good fit with a 
$\chi^2\sim107.6$ with 86 d.o.f. ($\chi_{red}^2\sim1.25$). The best-fit values of the three free parameters
are $Z_0=0.36\pm0.03$, $a=0.32\pm0.07$, and $\gamma=0.25\pm0.57$. These values represent a significant negative trend of $Z$ with  
the radius and again no significant evolution with redshift. 
That the evolution with redshift is even smaller when the radial dependence is taken into account suggests 
that part of the claimed evidence of a negative evolution of the integrated metallicity can be ascribed to the decrease 
with radius in the measured abundance, for instance when the metal content is estimated in relatively smaller regions 
in the nearby systems.

Although the constraints on the abundance evolution with the redshift
are far from being stringent, this is the first time that the abundance measured in the ICM of galaxy clusters has been parametrized 
as a function of both the distance from the center, and the redshift. This new result should be directly
compared with the models of diffusion of metals \citep[e.g.][]{ettori05,loewenstein06,calura07,cora08} and the simulations
of chemical enrichment \citep[e.g][]{tornatore04,rebusco05,tornatore07,schindler08,fabjan10} in the ICM.

\section{Conclusions}

We have presented a spatially resolved XMM-Newton analysis of the X-ray spectra of 39 clusters of galaxies at $0.4<z<1.4$, 
covering a temperature range of $1.5\,\la kT\,\la 11$~keV. The main goal of this paper has been to study how the abundance
evolves with redshift not only by analyzing a single emission measure performed on the whole cluster 
\citep[as in e.g. BLS07; MAU08;][]{anderson09},
but also by spatially resolving the cluster emission in three different regions: the core region (corresponding
to $0<r<0.15r_{500}$), the region immediately surrounding the core ($0.15r_{500}<r<0.4r_{500}$), and the outskirts of the cluster
($r>0.4r_{500}$).
This represents the first time that a spatially resolved analysis of the metal abundance evolution through cosmic time has
been attempted.

The main results of our analysis can be summarized as follows:
\begin{itemize}
\item Across the studied redshift range, we have not found any statistically significant abundance evolution with redshift 
considering the integrated emission from the whole cluster both at $r<0.6r_{500}$ and at $r=0.15-0.6r_{500}$.
The most significant deviation from no evolution was found in the $0-0.6r_{500}$ sample, where a power-law fit to the 
$z-Z$ distribution gives a $\gamma=0.75_{-0.74}^{+0.80}$ at a 90\% confidence level.
\item Dividing the emission into three radial bins, we could not find any significant evidence of abundance evolution (always below 
$1\sigma$). The main reasons
for this are probably the smaller number of counts (and the associated larger errors in $Z$) and smaller statistical samples for two
of the three radial bins.
\item We have not found any significant correlation between the abundance and temperature similar to that seen in other samples 
of galaxy clusters observed with other X-ray missions such as Chandra (BLS07) and ASCA \citep[e.g.][]{baumgartner05}. Using different
statistical methods (Spearman's rank, power-law fit, K-S test), the significance of a $kT-Z$ relation is always less than $1\sigma$. 
\item We compared our results for the single-emission-measure abundances with previous works in the literature. Our results are
fully consistent with the BLS07, MAU08, and \citet{anderson09} samples at $z>0.4$, as indicated by a power-law fit to the 
$z$-$Z$ distributions. The differences in the power-law slope $\gamma$ and in the normalizations at $z=0.6$ are always smaller 
than 2$\sigma$, with the largest discrepancy found when comparing with the MAU08 $0-1r_{500}$ sample.
\item We found that the presence or not of a single cluster (CLJ1415.1+3612) in the MAU08 sample could dramatically change the 
abundance evolution measured at $z>0.4$ from a power-law fit, such that in the $0.15-1r_{500}$ sample, 
$\gamma$ varies from $3.87_{-0.92}^{+1.06}$ (indicating strong
negative evolution) to $-0.15_{-1.23}^{+1.39}$ (no significant evolution). The impact that a single deviant 
object could have on the measured abundance evolution is a caveat about being careful in interpreting results based 
on such small samples.
\item We computed error-weighted means of the spatially resolved abundances in three redshift bins: $0.4<z<0.5$, $0.5<z<0.7$, and 
$z>0.7$. The abundance weighted-mean is consistent with being constant with redshift at every radius, given the large 1$\sigma$ error 
bars. A qualitative comparison with MAU08 and the predictions of the model of \citet{ettori05} shows how the trend in the 
error-weighted mean abundance with redshift 
in the 0.15-0.4$r_{500}$ radial bin complements nicely the measures of MAU08 (at $0.15r_{500}<r<r_{500}$), and broadly agrees with the 
predictions of \citet{ettori05} model. 
\item We quantified the dependence of the metal abundance on both the radius $r$ and the redshift $z$ by fitting  
the data points available from the spatially resolved analysis with a function of two variables 
$Z(r,z)=Z_0(1+(r/0.15r_{500})^2)^{-a}(z/0.6)^{-\gamma}$, with best-fit model values of $Z_0=0.36\pm0.03$, 
$a=0.32\pm0.07$, and $\gamma=0.25\pm0.57$. This relation depicts a significant negative trend of $Z$ with the radius 
and, although no significant evolution with the redshift is detected, it represents the first time that the abundance measured 
in ICM has been parametrized as a function of both $r$ and $z$ in a way that could be directly compared with the models 
of diffusion of metals \citep[e.g.][]{loewenstein06,calura07} and the simulations of chemical enrichment 
\citep[e.g][]{schindler08,fabjan10} in the ICM.
\end{itemize}

In general, the low quality of the statistics (and large errors in $Z$) associated with most of the clusters in the sample 
has prevented us fron drawing any statistically significant conclusion about the different evolutionary path the abundance may have 
traversed through cosmic time in the three different cluster regions considered.

We stress that the results presented in this paper (from both our new analysis of the XMM-Newton exposures
available at $z>0.4$ and those already presented in the literature) 
represent the most robust limits that could be derived from the existing XMM-Newton and Chandra archival data
on the evolution of the metal content of the ICM in massive halos. 
Larger samples of high redshift clusters observed in the X-rays and deeper observations of the clusters already 
present in both Chandra and XMM-Newton archive would be crucial to place statistically significant constraints on the 
evolution of metal abundances in galaxy clusters and to provide a robust modelling of the physical processes
responsible for the enrichment of the cluster plasma during its assembly over cosmic time.

\begin{acknowledgements}
We acknowledge financial contribution from contracts ASI-INAF I/023/05/0 and I/088/06/0.
FG acknowledges financial support from contract ASI-INAF I/009/10/0.
We thank the anonymous referee for the useful comments and suggestions that helped
to improve the presentation of the results in this paper.
\end{acknowledgements}

\bibliographystyle{aa}
\bibliography{alesbib.bib}

\begin{thebibliography}{41}
\expandafter\ifx\csname natexlab\endcsname\relax\def\natexlab#1{#1}\fi

\bibitem[{{Anders} \& {Grevesse}(1989)}]{anders89}
{Anders}, E. \& {Grevesse}, N. 1989, \gca, 53, 197

\bibitem[{{Anderson} {et~al.}(2009){Anderson}, {Bregman}, {Butler}, \&
  {Mullis}}]{anderson09}
{Anderson}, M.~E., {Bregman}, J.~N., {Butler}, S.~C., \& {Mullis}, C.~R. 2009,
  \apj, 698, 317

\bibitem[{{Andreon}(2010)}]{andreon10}
{Andreon}, S. 2010, \mnras, 407, 263

\bibitem[{{Arnaud}(1996)}]{arnaud96}
{Arnaud}, K.~A. 1996, in Astronomical Society of the Pacific Conference Series,
  Vol. 101, Astronomical Data Analysis Software and Systems V, ed. G.~H.
  {Jacoby} \& J.~{Barnes}, 17--+

\bibitem[{{Asplund} {et~al.}(2009){Asplund}, {Grevesse}, {Sauval}, \&
  {Scott}}]{asplund09}
{Asplund}, M., {Grevesse}, N., {Sauval}, A.~J., \& {Scott}, P. 2009, \araa, 47,
  481

\bibitem[{{Baldi} {et~al.}(2007){Baldi}, {Ettori}, {Mazzotta}, {Tozzi}, \&
  {Borgani}}]{baldi07}
{Baldi}, A., {Ettori}, S., {Mazzotta}, P., {Tozzi}, P., \& {Borgani}, S. 2007,
  \apj, 666, 835

\bibitem[{{Balestra} {et~al.}(2007){Balestra}, {Tozzi}, {Ettori}, {Rosati},
  {Borgani}, {Mainieri}, {Norman}, \& {Viola}}]{balestra07}
{Balestra}, I., {Tozzi}, P., {Ettori}, S., {et~al.} 2007, \aap, 462, 429
  (BLS07)

\bibitem[{{Baumgartner} {et~al.}(2005){Baumgartner}, {Loewenstein}, {Horner},
  \& {Mushotzky}}]{baumgartner05}
{Baumgartner}, W.~H., {Loewenstein}, M., {Horner}, D.~J., \& {Mushotzky}, R.~F.
  2005, \apj, 620, 680

\bibitem[{{Borgani} {et~al.}(2008){Borgani}, {Fabjan}, {Tornatore},
  {Schindler}, {Dolag}, \& {Diaferio}}]{borgani08}
{Borgani}, S., {Fabjan}, D., {Tornatore}, L., {et~al.} 2008, \ssr, 134, 379

\bibitem[{{Calura} {et~al.}(2007){Calura}, {Matteucci}, \& {Tozzi}}]{calura07}
{Calura}, F., {Matteucci}, F., \& {Tozzi}, P. 2007, \mnras, 378, L11

\bibitem[{{Coles} \& {Lucchin}(1995)}]{coles95}
{Coles}, P. \& {Lucchin}, F. 1995, {Cosmology. The origin and evolution of
  cosmic structure}, ed. {Coles, P.~\& Lucchin, F.}

\bibitem[{{Cora} {et~al.}(2008){Cora}, {Tornatore}, {Tozzi}, \&
  {Dolag}}]{cora08}
{Cora}, S.~A., {Tornatore}, L., {Tozzi}, P., \& {Dolag}, K. 2008, \mnras, 386,
  96

\bibitem[{{De Grandi} \& {Molendi}(2001)}]{degrandi01}
{De Grandi}, S. \& {Molendi}, S. 2001, \apj, 551, 153

\bibitem[{{Ettori}(2005)}]{ettori05}
{Ettori}, S. 2005, \mnras, 362, 110

\bibitem[{{Ettori} {et~al.}(2004){Ettori}, {Tozzi}, {Borgani}, \&
  {Rosati}}]{ettori04}
{Ettori}, S., {Tozzi}, P., {Borgani}, S., \& {Rosati}, P. 2004, \aap, 417, 13

\bibitem[{{Fabjan} {et~al.}(2010){Fabjan}, {Borgani}, {Tornatore}, {Saro},
  {Murante}, \& {Dolag}}]{fabjan10}
{Fabjan}, D., {Borgani}, S., {Tornatore}, L., {et~al.} 2010, \mnras, 401, 1670

\bibitem[{{Finoguenov} {et~al.}(2000){Finoguenov}, {David}, \&
  {Ponman}}]{finoguenov00a}
{Finoguenov}, A., {David}, L.~P., \& {Ponman}, T.~J. 2000, \apj, 544, 188

\bibitem[{{Irwin} \& {Bregman}(2001)}]{irwin01}
{Irwin}, J.~A. \& {Bregman}, J.~N. 2001, \apj, 546, 150

\bibitem[{{Kalberla} {et~al.}(2005){Kalberla}, {Burton}, {Hartmann}, {Arnal},
  {Bajaja}, {Morras}, \& {P{\"o}ppel}}]{kalberla05}
{Kalberla}, P.~M.~W., {Burton}, W.~B., {Hartmann}, D., {et~al.} 2005, \aap,
  440, 775

\bibitem[{{Leccardi} \& {Molendi}(2008{\natexlab{a}})}]{leccardi08b}
{Leccardi}, A. \& {Molendi}, S. 2008{\natexlab{a}}, \aap, 487, 461

\bibitem[{{Leccardi} \& {Molendi}(2008{\natexlab{b}})}]{leccardi08a}
{Leccardi}, A. \& {Molendi}, S. 2008{\natexlab{b}}, \aap, 486, 359

\bibitem[{{Liedahl} {et~al.}(1995){Liedahl}, {Osterheld}, \&
  {Goldstein}}]{liedahl95}
{Liedahl}, D.~A., {Osterheld}, A.~L., \& {Goldstein}, W.~H. 1995, \apjl, 438,
  L115

\bibitem[{{Loewenstein}(2006)}]{loewenstein06}
{Loewenstein}, M. 2006, \apj, 648, 230

\bibitem[{{Matteucci} \& {Vettolani}(1988)}]{matteucci88}
{Matteucci}, F. \& {Vettolani}, G. 1988, \aap, 202, 21

\bibitem[{{Maughan} {et~al.}(2008){Maughan}, {Jones}, {Forman}, \& {Van
  Speybroeck}}]{maughan08}
{Maughan}, B.~J., {Jones}, C., {Forman}, W., \& {Van Speybroeck}, L. 2008,
  \apjs, 174, 117 (MAU08)

\bibitem[{{Nousek} \& {Shue}(1989)}]{nousek89}
{Nousek}, J.~A. \& {Shue}, D.~R. 1989, \apj, 342, 1207

\bibitem[{{Peacock}(1999)}]{peacock99}
{Peacock}, J.~A. 1999, {Cosmological Physics}, ed. {Peacock, J.~A.}

\bibitem[{{Peebles}(1993)}]{peebles93}
{Peebles}, P.~J.~E. 1993, {Principles of Physical Cosmology}, ed. {Peebles,
  P.~J.~E.}

\bibitem[{{Rebusco} {et~al.}(2005){Rebusco}, {Churazov}, {B{\"o}hringer}, \&
  {Forman}}]{rebusco05}
{Rebusco}, P., {Churazov}, E., {B{\"o}hringer}, H., \& {Forman}, W. 2005,
  \mnras, 359, 1041

\bibitem[{{Renzini}(1997)}]{renzini97}
{Renzini}, A. 1997, \apj, 488, 35

\bibitem[{{Rosati} {et~al.}(2002){Rosati}, {Borgani}, \& {Norman}}]{rosati02}
{Rosati}, P., {Borgani}, S., \& {Norman}, C. 2002, \araa, 40, 539

\bibitem[{{Schindler} \& {Diaferio}(2008)}]{schindler08}
{Schindler}, S. \& {Diaferio}, A. 2008, \ssr, 134, 363

\bibitem[{{Snowden} {et~al.}(2004){Snowden}, {Collier}, \& {Kuntz}}]{snowden04}
{Snowden}, S.~L., {Collier}, M.~R., \& {Kuntz}, K.~D. 2004, \apj, 610, 1182

\bibitem[{{Snowden} {et~al.}(2008){Snowden}, {Mushotzky}, {Kuntz}, \&
  {Davis}}]{snowden08}
{Snowden}, S.~L., {Mushotzky}, R.~F., {Kuntz}, K.~D., \& {Davis}, D.~S. 2008,
  \aap, 478, 615

\bibitem[{{Tamura} {et~al.}(2004){Tamura}, {Kaastra}, {den Herder}, {Bleeker},
  \& {Peterson}}]{tamura04}
{Tamura}, T., {Kaastra}, J.~S., {den Herder}, J.~W.~A., {Bleeker}, J.~A.~M., \&
  {Peterson}, J.~R. 2004, \aap, 420, 135

\bibitem[{{Tornatore} {et~al.}(2007){Tornatore}, {Borgani}, {Dolag}, \&
  {Matteucci}}]{tornatore07}
{Tornatore}, L., {Borgani}, S., {Dolag}, K., \& {Matteucci}, F. 2007, \mnras,
  382, 1050

\bibitem[{{Tornatore} {et~al.}(2004){Tornatore}, {Borgani}, {Matteucci},
  {Recchi}, \& {Tozzi}}]{tornatore04}
{Tornatore}, L., {Borgani}, S., {Matteucci}, F., {Recchi}, S., \& {Tozzi}, P.
  2004, \mnras, 349, L19

\bibitem[{{Vikhlinin}(2006)}]{vikhlinin06}
{Vikhlinin}, A. 2006, \apj, 640, 710

\bibitem[{{Vikhlinin} {et~al.}(2007){Vikhlinin}, {Burenin}, {Forman}, {Jones},
  {Hornstrup}, {Murray}, \& {Quintana}}]{vikhlinin07}
{Vikhlinin}, A., {Burenin}, R., {Forman}, W.~R., {et~al.} 2007, in Heating
  versus Cooling in Galaxies and Clusters of Galaxies, ed. {H.~B{\"o}hringer,
  G.~W.~Pratt, A.~Finoguenov, \& P.~Schuecker }, 48--+

\bibitem[{{Vikhlinin} {et~al.}(2005){Vikhlinin}, {Markevitch}, {Murray},
  {Jones}, {Forman}, \& {Van Speybroeck}}]{vikhlinin05}
{Vikhlinin}, A., {Markevitch}, M., {Murray}, S.~S., {et~al.} 2005, \apj, 628,
  655

\bibitem[{{Voit}(2005)}]{voit05}
{Voit}, G.~M. 2005, Reviews of Modern Physics, 77, 207

\end{thebibliography}

\end{document}